\documentclass[a4paper,11pt]{article}
\pdfoutput=1
\usepackage{jheppub}

\usepackage{epsf}
\usepackage{epsfig}
\usepackage{subfigure}
\usepackage{mathtools}
\usepackage{hhline}
\usepackage{float}
\restylefloat{table}
\usepackage{multirow}
\usepackage{nicefrac}
\usepackage{epstopdf}
\usepackage{url}
\usepackage{multicol}
\usepackage{array}
\usepackage[normalem]{ulem}
\usepackage{color, colortbl}

\usepackage[compat=1.0.0]{tikz-feynman}

\allowdisplaybreaks

\newcommand{\be}{\begin{eqnarray}}
\newcommand{\ee}{\end{eqnarray}}

\newcommand{\ba}{\begin{array}}
\newcommand{\ea}{\end{array}}
\newcommand{\bee}{\begin{equation}\ba{c}}
\newcommand{\eee}{\ea\end{equation}}

\newcommand{\bi}{\begin{itemize}}
\newcommand{\ei}{\end{itemize}}

\definecolor{tableblue}{rgb}{0.2, 0.2, 0.8}
\definecolor{tablered}{rgb}{0.8, 0.2, 0.2}

\toccontinuoustrue

\title{Isolating New Physics signatures of vectorlike quarks and heavy Higgses over multi $b$-jet backgrounds}

\author{Enrico Lunghi$^1$,}
\author{Beni Pazar$^1$}
\affiliation{
$^1$Physics Department, Indiana University, Bloomington, IN 47405, USA \\
}
\emailAdd{bpazar@iu.edu}
\emailAdd{elunghi@iu.edu}

\abstract{We discuss models in which vectorlike quarks and a second Higgs doublet are simultaneously present and couple to the third generation of SM quarks. A general feature of these models is that cascade decays of vectorlike fermions into heavy Higgses (or vice versa, depending on their masses) dominate their respective branching ratios. The resulting collider signals involve a large multiplicity of $b$-jets and offer formidable experimental challenges. We present a detailed and realistic analysis of the expected sensitivity of dedicated searches at ATLAS and CMS.
}

\begin{document}

\maketitle

\section{Introduction}
\label{sec:introduction}

Since the discovery of a single light neutral Higgs~\cite{ATLAS:2012yve, CMS:2012qbp}, the search for additional Higgs bosons has been a major focus of the LHC experimental program. One of the simplest extensions to the Standard Model (SM) is the introduction of a  second Higgs doublet; the resulting theories are called Two Higgs Doublet Models (2HDM) and have been the subject of extensive studies.  In general, any 2HDM has five physical Higgs bosons, two CP-even ($h$ and $H$), one CP-odd ($A$), and two charged ($H^\pm$) Higgses and are classified based on the Yukawa couplings of the Higgs bosons to fermions. In the 2HDM Type-II model, which is the only model that can be embedded in the minimal supersymmetric extensions of the SM (MSSM)~\cite{Haber:1984rc, Martin:1997ns}, the Higgs doublet that gives mass to up-type quarks also gives mass to the charged leptons. The other doublet gives mass to down-type quarks and neutrinos. Note that while this coupling structure appears naturally in the MSSM, it has to be imposed in non-supersymmetric new physics scenarios. We refer, for instance, to ref.~\cite{Branco:2011iw} for a detailed review of the various 2HDM variants that have been considered in the literature.

Another common extension of the SM is the addition of fermions whose chiral components are assigned identical transformation properties under the SM gauge group. These fermions, which are known as vectorlike fermions, were originally considered because they tend to appear naturally in Grand Unified Theories~\cite{Ramond:1981zz, delAguila:1982fs, Hewett:1988xc} and models with Dynamical Electroweak Symmetry Breaking~\cite{Perelstein:2003wd, Contino:2006qr, Matsedonskyi:2012ym}. Vectorlike fermions, in general, couple to all Higgs fields present in the theory with Yukawa interactions which, generally, do not provide the main contributions to their own masses; in fact, the vectorlike nature of the charge assignments allows for the presence of arbitrarily large tree-level mass terms. All vectorlike fermions need to possess Yukawa couplings to at least one generation of SM fermions to allow for their decays to light matter (this, in turn, has implications for the flavor structure of the theory~\cite{Branco:1986my, Kaplan:1991dc, Branco:1992wr}). 

These two extensions of the SM have been the subject of extensive research and have been used to address a variety of theoretical issues. In the context of supersymmetric Grand Unified Theories, vectorlike fermions have been used to address gauge coupling unification~\cite{Babu:1996zv, Kolda:1996ea, Ghilencea:1997yr, AmelinoCamelia:1998tm, BasteroGil:1999dx, Dermisek:2012as, Dermisek:2012ke, Dermisek:2017ihj, Dermisek:2018hxq}, the impact on 
electroweak symmetry breaking and the Higgs boson mass~\cite{Babu:2008ge, Martin:2009bg, Dermisek:2016tzw}, and the impact on the weak mixing angle~\cite{Moroi:1993zj, Dermisek:2017ihj}. Aside from supersymmetric models, vectorlike fermions have been introduced to address various anomalies like the tension in precision Electroweak observables~\cite{Choudhury:2001hs, Dermisek:2011xu, Dermisek:2012qx, Batell:2012ca} and the muon g-2 anomaly~\cite{Kannike:2011ng, Dermisek:2013gta, Dermisek:2014cia}.

When both a second Higgs doublet and some vectorlike fermions are present, it is very natural for cascade decays of vectorlike fermions to Higgs (or vice versa, depending on the mass hierarchy) to have very large branching ratios. In this situation, standard searches loose power leading to situations in which new Higgses and fermions hide in plain sight.\footnote{Here we refer to searches in which vectorlike quarks are assumed to decay to $(W^\pm, Z,h)+\text{SM quark}$ and heavy Higgses are looked for in the usual channels ($t\bar t$, $b\bar b$, $\tau^+\tau^-$, $\gamma \gamma$, ...).}  A review of the currently allowed parameter space for these models and of the many possible signatures has been presented in refs.~\cite{Dermisek:2019vkc, Dermisek:2016via, Dermisek:2020gbr, Dermisek:2019heo}.

\begin{figure}
\begin{center}
\includegraphics[width=0.45 \linewidth]{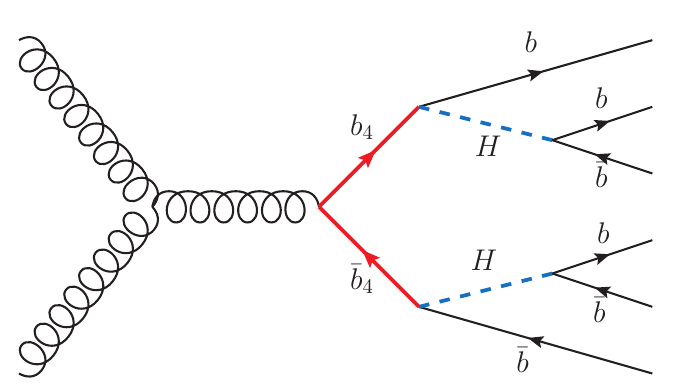}
\;\;
\includegraphics[width=0.45 \linewidth]{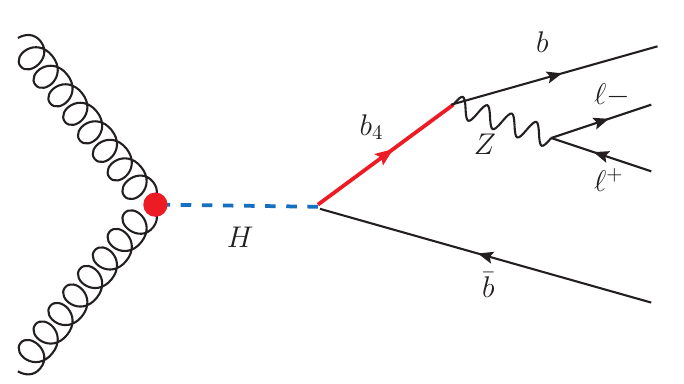}
\end{center}
\caption{The left and right panels display the Feynman diagrams for $pp \to b_4 \bar b_4 \to (Hb) (H \bar b) \to (b\bar b b) (b\bar b \bar b)$ and $pp\to H \to b_4 b \to \bar b b \ell^+\ell^-$, respectively.}
\label{fig:Feynman}
\end{figure}

In this paper we focus on certain decay modes which lead to final states possessing a large multiplicity of $b$-jets:
\begin{alignat}{4}
pp &\to b_4 \bar b_4      \to (Hb) (H \bar b) &&\to (b\bar b b) (b\bar b \bar b) \; , \\ 
pp &\to H \to b_4 \bar b \to  Z b  \bar b  &&\to \ell^+ \ell^- b \bar b \;\;\; (\ell = e,\mu) \; ,
\end{alignat}
where $b_4$ is the lightest vectorlike quark with charge $-1/3$ and $H$ is the neutral Heavy Higgs\footnote{An identical process involving the CP-odd Higgs $A$ is possible. If the two heavy neutral Higgses are close in mass, the net effect is a doubling of the signal cross section}. The Feynman diagrams for these processes are shown in figure~\ref{fig:Feynman}. In order to disentangle the very large QCD backgrounds (originating from light or charm jets mistagged as $b$-jets) we make use of a tagging strategy (which we refer to as ``1b2b-tagging''), based on ideas proposed in ref.~\cite{Goncalves:2015prv} and fully developed in ref.~\cite{Lunghi:2023vzx}, in which various jet substructure observables are used to distinguish jets initiated by partonic gluons and b-quarks. These two signals have been investigated previously in refs.~\cite{Dermisek:2020gbr, Dermisek:2019heo} where the impact of a possible 1b2b-tagger has been ball-parked based on the results presented in ref.~\cite{Goncalves:2015prv}. Recently several ATLAS groups have expressed interest in performing these searches; thus motivating a detailed and realistic reanalysis of the expected sensitivities of these processes. 

The paper is organized as follows. In section~\ref{sec:Model} we discuss the 2HDM Type-II model with the addition of vectorlike quarks. In section~\ref{sec:6b} and~\ref{sec:Zbb} we present the analysis strategy and expected reach  of cascade decays of vectorlike quarks through Heavy Higgses and Heavy Higgses through vectorlike quarks, respectively. Finally in section~\ref{sec:Conclusion} we present our conclusions.

\section{Two Higgs Doublet Model in presence of Vectorlike Quarks}
\label{sec:Model}
The 2HDM model we consider possesses a $Z_2$ symmetry, under which the two Higgs doublets $H_u$ and $H_d$ have opposite parity, which forces a type-II structure on the Yukawa interactions . Moreover, we introduce both doublet ($Q_{L,R}$) and singlet ($T_{L.R}$ and $B_{L,R}$) vectorlike fermions which share the same quantum numbers as the SM quarks. The actual charge assignments that we consider are summarized in table~\ref{tab:Model}. The resulting Yukawa/Mass part of the Lagrangian is:
\begin{align}
 \begin{split}
-\mathcal{L} \supset &+y_d^{ij}\bar q_L^i d_R^j H_d + \lambda_B^i \bar q_L^i B_R H_d + \lambda_Q^j\bar Q_L d_R^j H_d + \lambda\bar Q_L B_R H_d + \bar\lambda H_d^\dag\bar B_L Q_R \\
    &+y_u^{ij}\bar q_L^i u_R^j H_u + \kappa_T^i\bar q_L^i T_R H_u + \kappa_Q^j\bar Q_L u_R^j H_u + \kappa\bar Q_L T_R H_u + \bar\kappa H_u^\dag \bar T_L Q_R \\
    &+M_Q \bar Q_L Q_R + M_T \bar T_L T_R + M_B \bar B_L B_R + h.c.
\end{split}
\label{eq:Lagrangian}
\end{align}
The Higgs potential (whose details we are not interested in) yields vacuum expectation values for the neutral components of the two Higgs doublets, $\langle H_u \rangle = (v_u,0)$ and $\langle H_u \rangle = (0,v_d)$, which we conventionally parameterize in terms of the standard vacuum expectation value $v = \sqrt{v^2_u+v^2_d}=174\; {\rm GeV}$ and $\tan{\beta}\equiv v_u/v_d$. The mass parameters $M_{Q,T,B}$ are the dominant contributions to the physical masses of the vectorlike quarks while the various Yukawa interactions ($\lambda_{B,Q}$, $\kappa_{T,Q}$) and the doublet/singlet mixing couplings ($\lambda$, $\bar \lambda$, $\kappa$, $\bar \kappa$) control the cascade decays involving heavy Higgses and vectorlike quarks.

\begin{table}[h]
  \begin{center}
  \begin{tabular}{ccccccccc}\hline\hline
     & $q^i_L$& $u^i_R$ & $d^i_R$ & $Q_{L,R}$ & $T_{L,R}$ & $B_{L,R}$ & $H_d$ & $H_u$ \\
    \hline
    $SU(2)_L$ & 2 & 1 & 1 & 2 & 1 & 1 & 2 & 2 \\ \hline
    $U(1)_Y$ & $\frac{1}{6} $& $\frac{2}{3} $& $-\frac{1}{3}$ & $\frac{1}{6} $& $\frac{2}{3} $& $-\frac{1}{3}$ & $\frac{1}{2}$& $-\frac{1}{2}$ \\ \hline
    $Z_2$ & + & + & - & + & + & - & - & + \\ \hline
  \end{tabular}
  \end{center}
  \caption{Quantum numbers of the SM quarks ($i = 1,2,3$), vectorlike quarks, and two Higgs doublets.}
  \label{tab:Model}
\end{table}

In Eq.~(\ref{eq:Lagrangian}) we allow only for couplings between vectorlike quarks and the third generation of SM fermions. It is in principle possible to consider couplings to more than one SM generation; in this case collider signatures would be harder to disentangle because of the presence of genuine light jets in the final state but there would be the possibility of sizable contributions to flavor changing transitions (e.g. if couplings to the second and third generations are simultaneously present, sizable contributions to the the Wilson coefficients relevant for inclusive and exclusive $b\to s \ell^+ \ell^-$ decays can be generated). 

In the first scenario that we consider, the heavy neutral CP-even Higgs is lighter than the lightest vectorlike quark and decays predominantly into $b\bar b$ at large $\tan\beta$ and $t\bar t$ at small $\tan\beta$ (see, for instance, figure~3 of ref.~\cite{Dermisek:2019heo}). The vectorlike quark branching ratio into heavy Higgses on the other hand, can easily be of order $O(1)$ as can be seen in figure~1 of ref.~\cite{Dermisek:2019vkc}. 

In the second scenario, in which a vectorlike quark is lighter than the heavy neutral Higgs, the branching ratio $H\to b_4 b$ can be close to 100\% (see figure~3 of ref.~\cite{Dermisek:2019heo}) while the vectorlike quarks decays to $W$, $Z$ and $h$ (Standard Model Higgs) final states. If the vectorlike quark mass is much larger than the electroweak scale, the singlet and doublet branching ratios follows the simple structure imposed by the Goldstone Boson Equivalence Theorem. For vectorlike quark singlet the partial widths into $(W,Z,h)$ are in 2:1:1 ratios, corresponding to (50\%, 25\%, 25\%) branching ratios, respectively. In the doublet case, the partial widths into $Z$ and $h$ are identical but the $W$ one is arbitrary (depending on the couplings $\kappa_Q$ and $\lambda_Q$). Therefore, the branching ratio $b_4 \to Z b$ can be in the range $[0,50\%]$.  

In the next two sections we present detailed studies of these two signatures. Simulated events for both signals and backgrounds have been generated using MadGraph5~\cite{Alwall:2011uj}, interfaced to Pythia8~\cite{Sjostrand:2006za, Sjostrand:2014zea} for parton shower and hadronization, and Delphes~\cite{deFavereau:2013fsa} for detector simulation. The model has been implemented using FeynRules~\cite{Alloul:2013bka} by S.~Shin and N.~McGinnis, has been used for the analysis presented in refs.~\cite{Dermisek:2019vkc, Dermisek:2019heo, Dermisek:2020gbr, Dermisek:2021zjd, Dermisek:2022xal}, and has been successfully incorporated in the ATLAS framework.

\section{$M_{\text{VLQ}} > M_H$: $pp \to b_4 \bar b_4 \to (Hb) (H \bar b) \to (b\bar b b) (b\bar b \bar b)$}
\label{sec:6b}

We start be considering the $M_{b_4} > M_H$ mass hierarchy. In this situation the pair production cross section of vectorlike quarks depends exclusively on their mass and can be calculated using standard codes (see, for instance, figure~2 of ref.~\cite{Dermisek:2021zjd} where a combination of the MadGraph5~\cite{Alwall:2011uj} and Top++~\cite{Czakon:2011xx} have been used). 

The branching ratio for the subsequent decay $b_4 \to H b$ can then be easily close to 100\% (see figure~1 of ref.~\cite{Dermisek:2019vkc}). The reason for this is that the Yukawa couplings $\lambda_B$ and $\lambda_Q$ introduced in Eq.~(\ref{eq:Lagrangian}) yield partial widths of vectorlike quarks into $W^\pm$, $Z$ and $h$ which vanish at large $\tan\beta$ (see table~2 of ref.~\cite{Dermisek:2019vkc}); thus, assuming that these Yukawa couplings are not too small, the branching ratio into heavy neutral Higgs is necessarily very large. Additionally, taking into account that the branching ratio ${\rm BR} (H\to b\bar b)$ dominates at large $\tan\beta$ (see figure~3 of ref.~\cite{Dermisek:2019heo}), we see that in a vast region of parameter space (sizable Yukawa couplings between vectorlike and SM fermions and large $\tan\beta$) the branching ratio $b_4 \to H b \to b\bar b b$ can be close to unity (the upper limit is about 90\% due to unavoidable $H\to \tau\tau$ decays).

For each vectorlike quark mass, we consider two mass configurations: $M_H = 1\; \text{TeV}$ and $M_H = M_{b_4} - 200\; {\rm GeV}$ (close to the kinematic threshold). In general, the presence of a larger gap between the $b_4$ and $H$ masses results in harder $b$-jets which are easier to detect over multi-jet backgrounds; thus, we expect sensitivities in the $M_H = 1\; \text{TeV}$ scenarios to be stronger.

In the analysis strategy that we pursue, we construct a signal by requesting four well separated $b$-tagged jets at large $p_T$. We simulate background events in the $pp \to (4j, 2b 2j, 4b)$ channels, where $j$ represent a light parton. We note that an actual implementation of this analysis at ATLAS or CMS would require a data-driven approach to the background estimation (see, for instance, refs.~\cite{ATLAS:2023mny, CMS:2021yzl, ATLAS:2017ble}).

\begin{table}[t]
  \begin{center}
  \begin{tabular}{|c|c||c|c|c|c|c|c||c|c|}\hline
   $M_{b_4}$(TeV) & $M_H$(TeV) & $H_{T_b}$ & $p_{T_0}$ & $p_{T_1}$ & $p_{T_2}$ & $p_{T_3}$ & $M_{j_1 j_2 j_3}$ & \multicolumn{2}{c|}{$\text{BR}_{\rm lim}$}  \\ 
&&&&&&& &   $3\; \text{ab}^{-1}$ & $140 \; \text{fb}^{-1}$ \\
    \hline\hline
    1.5 & 1.0 & 1450 & 550 & 250 & 250 & 250 & 500 & 0.236 & 0.634\\
    1.5 & 1.3 & 1450 & 550 & 150 & 100 & 50 & 700 & 0.287 & 0.679\\
    \hline
    1.8 & 1.0 & 1450 & 550 & 250 & 250 & 250 & 800 & 0.475 & 1.289\\
    1.8 & 1.6 & 1450 & 550 & 150 & 100 & 50 & 750 & 0.673 & 1.596\\
    \hline
    2.0 & 1.0 & 1450 & 550 & 250 & 250 & 250 & 900 & 0.744 & 2.029\\
    2.0 & 1.8 & 1450 & 550 & 150 & 100 & 50 & 800 & 1.09& 2.595\\
    \hline
    2.5 & 1.0 & 1450 & 550 & 250 & 250 & 250 & 1000 & 2.361 & 6.515\\
    2.5 & 2.3 & 1450 & 550 & 150 & 100 & 50 & 800 & 3.749 & 8.923\\
    \hline
  \end{tabular}
  \end{center}
  \caption{$pp  \to 6b$. Cuts (in GeV) applied to each mass configuration studied and corresponding expected upper bounds on the $b_4 \to H b \to b\bar b b$ branching ratios corresponding to integrated luminosities equal to $3\; \text{ab}^{-1}$ and $140\; \text{fb}^{-1}$. The $H_{T_b}$ and $p_{T_i}$ cuts are identical for both values of the heavy Higgs masses, $M_H=1\; \text{TeV}$ and $M_H=M_{b_4}-200\; \text{GeV}$.}
\label{tab:6bCuts}
\end{table}
\begin{figure}[t]
  \begin{center}
    \includegraphics[width=0.99 \linewidth]{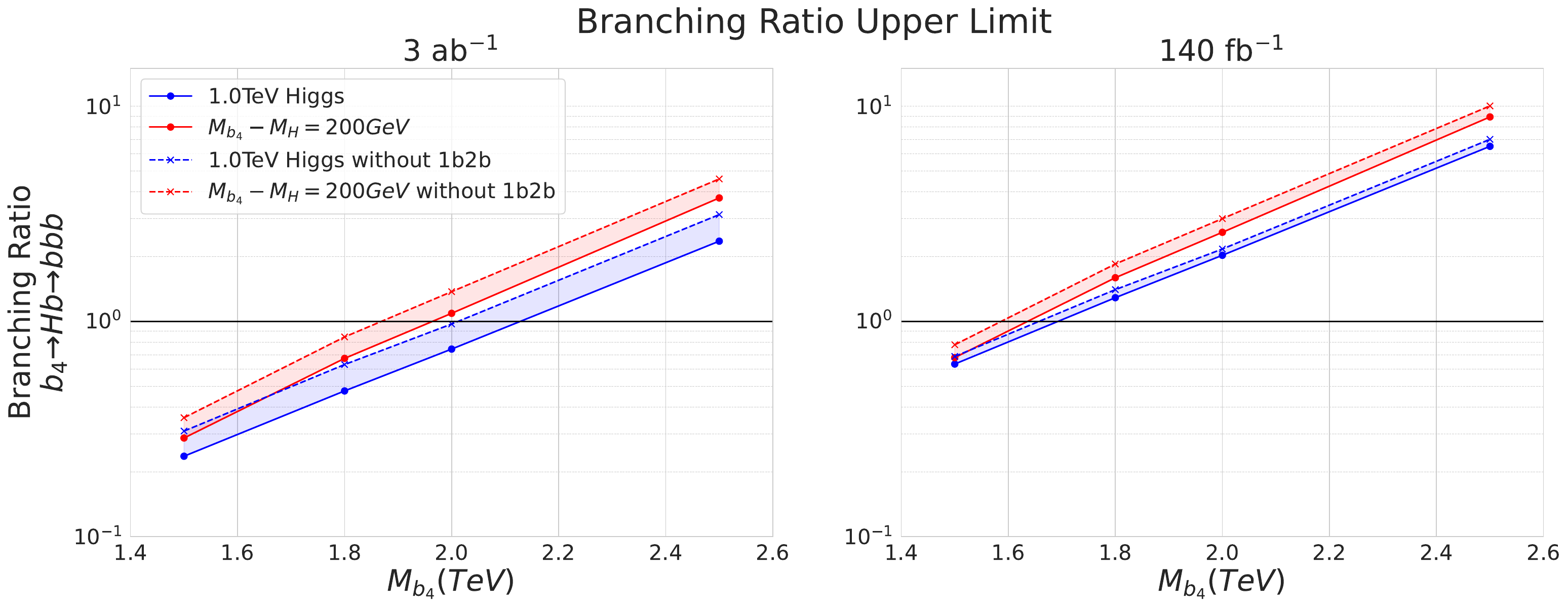}
  \end{center}
  \caption{$pp  \to 6b$: upper bounds on the $b_4\to Hb\to b\bar bb$ branching ratio. The dashed lines correspond removing the 1b2b tagger requirement from the analysis.}
  \label{fig:BRup6b}
\end{figure}

The variables we consider are the total transverse momentum of $b$-jets ($H_{T_b}$) which we use instead of the more common scalar sum of jet transverse momenta ($H_T$),  the transverse momentum of the leading four $b$-tagged jets ($p_{T_i}$, $i\in [0,1,2,3]$), and the invariant mass of the leading three jets. Note that which combination of three jets' invariant mass peaks the most near $M_{b_4}$ depends on the actual $b_4$ and $H$ masses. The main rationale for preferring $H_{T_b}$ over the more conventional $H_T$ is that we combine a hard cut on this variable coupled with a quite mild cut on the $p_T$ of the softest $b$-jet: $p_T^{\rm cut} (b) \lesssim H_{T_b}^{\rm cut}/N_{\rm jets}$. Under these conditions, a cut on $H_T$ would be saturated by a large multiplicity of non $b$-tagged and low $p_T$ jets rather than by a small number of hard $b$-jets. Thus, the use of $H_T$ would provide less discriminatory power against the signal we look for and create difficulties in the Monte Carlo simulation.

For each event, we require at least four b-tagged jets (which need not to be the four hardest jets in the event) with $|\eta| < 3$  and $\Delta R$\footnote{$\Delta R$ between two 4-momenta is defined as $\sqrt{(\Delta \eta)^2 + (\Delta \phi)^2}$, where $\eta$ and $\phi$ are pseudorapidity and azimuthal angle of the two 4-momenta, respectively.} between any two jets larger than 0.5. We then require the presence of at least four jets which are both b- and 1b2b-tagged. Given the enormous statistics that would be needed in actually imposing these taggers exactly on the multi-jet background, we apply both taggers by reweighing the events: the b-tagging formula is taken from a fit to the CMS tagging algorithm as implemented in Delphes, while the 1b2b-tagging efficiencies are taken from the results we published in ref.~\cite{Lunghi:2023vzx} (adopting the 80\% signal efficiency working point).

We vary the possible cuts over a wide range of possibilities and choose the configuration which maximizes the sensitivity. The optimal cuts we find are summarized in table~\ref{tab:6bCuts} for various $b_4$ and $H$ mass configurations. In the table we present also the expected upper limits on the branching ratio $\text{BR}(b_4 \to H b \to b\bar b)$ calculated for total integrated luminosities equal to $140\; \text{fb}^{-1}$ and $3 \; \text{ab}^{-1}$. In figure~\ref{fig:BRup6b} we present the the upper limit on the branching ratio as a function of the vectorlike quark and Higgs masses; additionally we show how the upper bounds change if no use is made of the 1b2b-tagger. 

These results show that an actual implementation of the 1b2b-tagger yields results that are actually very similar to those presented in ref.~\cite{Dermisek:2020gbr} where the tagger impact had been estimated using combinatoric arguments based on the early work presented in ref.~\cite{Goncalves:2015prv}. In this more realistic study we find sensitivity up to vectorlike bottom mass near 2.1 TeV.

Additional distributions are presented in figures~\ref{fig:HTb1}, \ref{fig:HTb2}, \ref{fig:MJJ1}, \ref{fig:MJJ2}, \ref{fig:MJJJ1} and \ref{fig:MJJJ2}. 

\section{$ M_H > M_{\text{VLQ}}$: $pp\to H\to b_4 \bar b\to (Zb)\bar b\to \ell^+\ell^- b\bar b$}
\label{sec:Zbb}

We now consider the inverted hierarchy for which  $M_H > M_{b_4}$. In this case, the production cross section is controlled by $pp\to H$ and is sensitive not only to the heavy Higgs mass but also to details of the heavy Higgs couplings to quarks (in particular, the main relevant parameter is $\tan\beta$). We begin with the extraction of model independent expected upper limits on the signal cross section which depend only on the vectorlike quark and Higgs masses. Besides leptons and jets transverse momentum, various invariant masses and the total transverse energy into $b$-jets ($H_{T_b}$) we also consider the vectorial sum of the charged leptons transverse momenta (i.e. the transverse momentum of the $Z$ boson) which is expected to be large for charged leptons produced in the decay of a heavy particle.  

\begin{table}[t]
  \begin{center}
  \begin{tabular}{|c|c||c|c|c|c|c|c|c|c|c|}\hline
    $M_H$(TeV) & $M_{b_4}$(TeV)  & $P_{T_{ll}}$ & $P_{T_{l_1}}$ & $H_{T_b}$ & $P_{T_{j_1}}$  & $M_{llj_0}$ & $M_{llj_1}$ & $M_{lljj}$ \\
    \hline\hline
    1.5 & 1.0 & 300 & 50 & 550 & 250  & 200 & 200 & 1200 \\ \hline
    1.5 & 1.3 & 300 & 50 & 450 & 50  & 1100 & 150 & 1250 \\ \hline
    2.0 & 1.0 & 300 & 50 & 800 & 150 & 400 & 700 & 1550 \\ \hline
    2.0 & 1.5 & 300 & 50 & 600 & 200 & 1200 & 250 & 1700 \\ \hline
    2.0 & 1.8 & 300 & 150 & 450 & 50 & 1500 & 150 & 1700 \\ \hline
    2.5 & 1.0 & 300 & 50 & 750 & 100  & 750 & 700 & 1900 \\ \hline
    2.5 & 1.5 & 300 & 50 & 750 & 300  & 1100 & 550 & 1950 \\ \hline
    2.5 & 1.8 & 300 & 150 & 700 & 150  & 1150 & 350 & 2100 \\ \hline
    2.5 & 2.3 & 550 & 150 & 300 & 50  & 2000 & 150 & 2150 \\ \hline
  \end{tabular}
  \end{center}
  \caption{$pp\to  2 \ell 2 b$: optimal cuts (in GeV) for various mass configurations.}  \label{tab:ZbbCuts}
\end{table}

As a starting point we generated signal and background events with the following basic cuts: $|\eta_\ell|\leq 2.4$, $M_{\ell^+ \ell^-}\in [66,116]$, ${p_T}_{\ell\ell} > 300\; \text{GeV}$, ${p_T}_{\ell_0} > 150\; \text{GeV}$, ${p_T}_{\ell_1} > 50\; \text{GeV}$, $|\eta_j|\leq 3$, $\Delta_R(j_0,j_1) > 0.5$, ${p_T}_{j_0} > 200\; \text{GeV}$ and ${p_T}_{j_1} > 100\; \text{GeV}$.

\begin{figure}[t]
  \begin{center}
    \includegraphics[width=0.49 \linewidth]{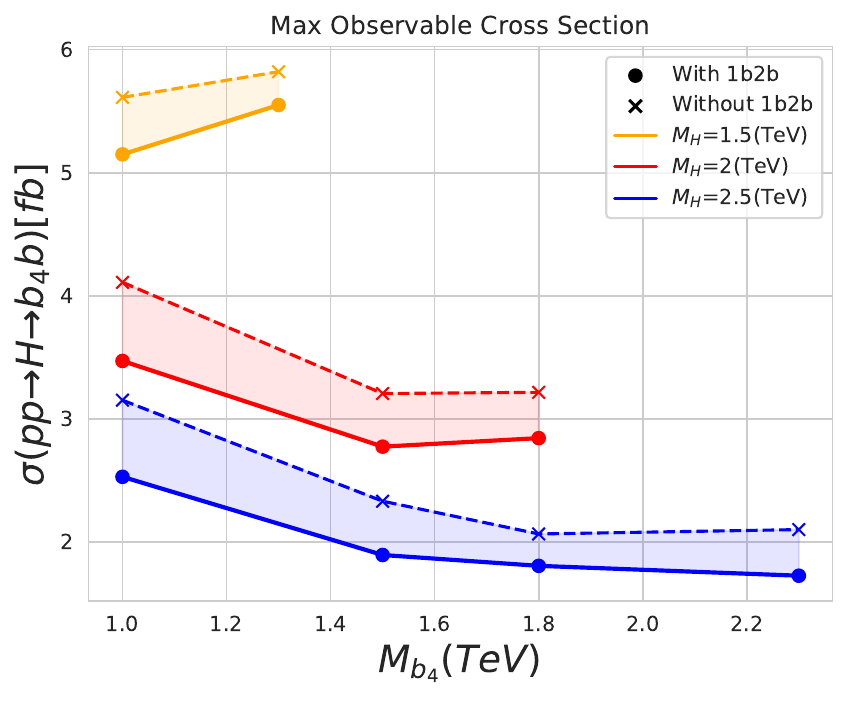}
    \includegraphics[width=0.49 \linewidth]{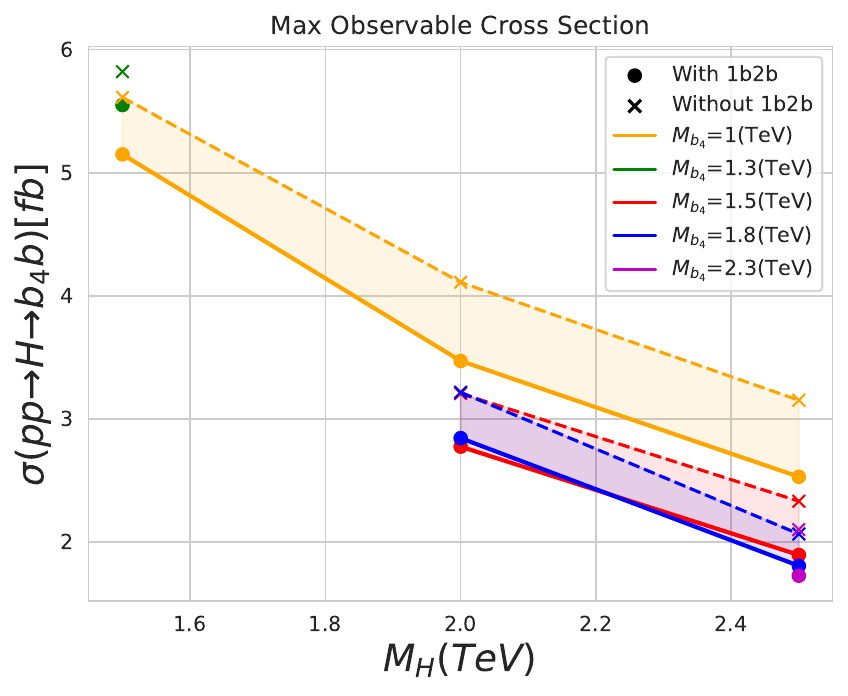}
  \end{center}
  \caption{$pp\to 2 \ell 2  b $: expected upper bounds on $\sigma(pp\to H \to b_4 \bar b)$ signal cross section for various mass configurations with and without the application of the 1b2b tagger.}
  \label{fig:CSupZbb}
\end{figure}

\begin{figure}[t]
  \begin{center}
    \includegraphics[width=0.92 \linewidth]{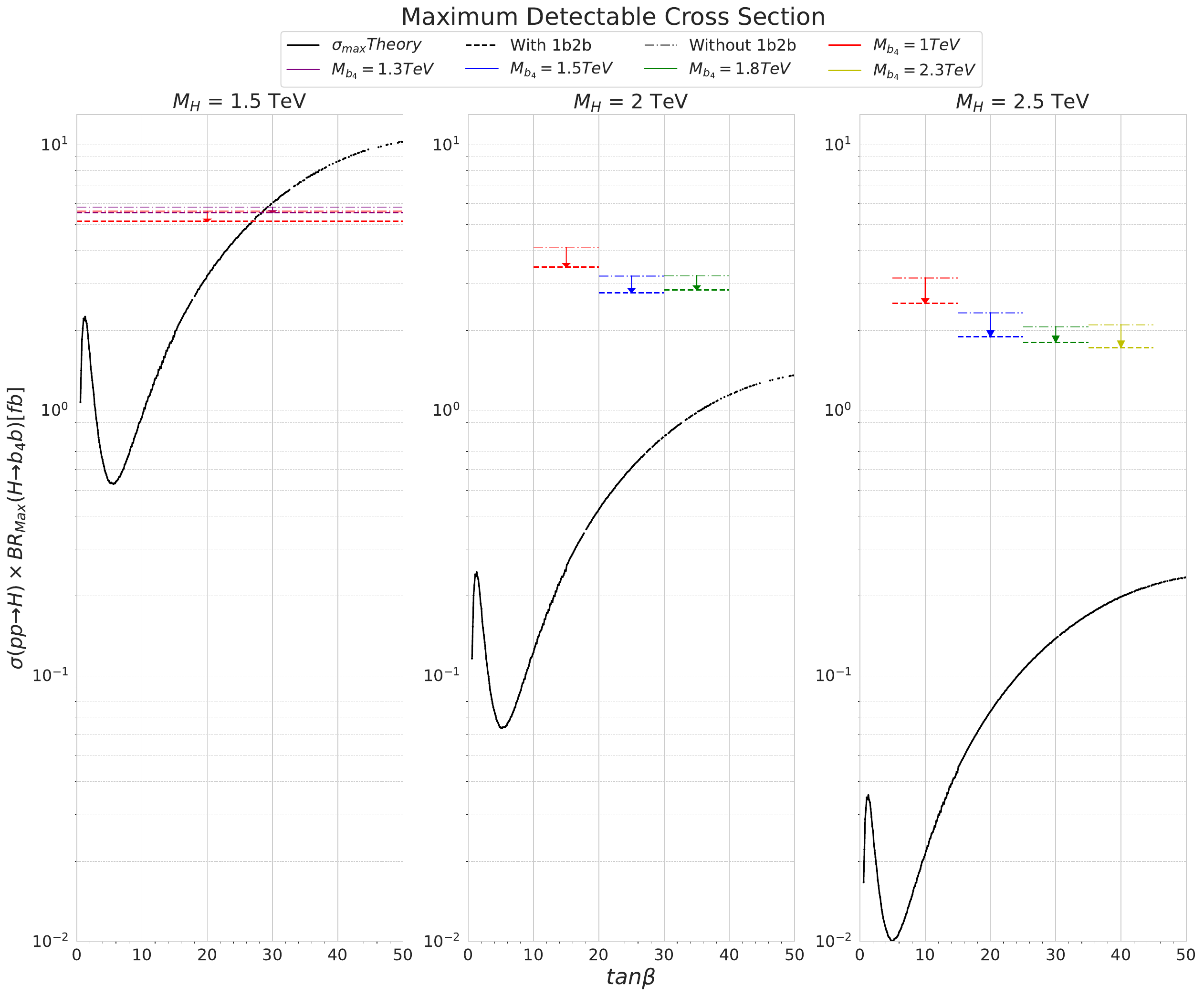}
  \end{center}
  \caption{$pp\to 2 \ell 2 b$: upper bounds on Cross Section for different values of $M_H$. For the $M_H = 2$ and 2.5 TeV cases we truncated the horizontal lines for ease of reading.}
  \label{fig:CSupZbb_Theory}
\end{figure}

\begin{table}[t]
  \begin{center}
  \begin{tabular}{|c|c||c|c|c|c|c|}\hline
    $M_H$ & $M_{b_4}$  & Signal & $\sigma_{jj}^{\rm bkgd}$(ab) & $\sigma_{b\bar b}^{\rm bkgd}$(ab) &  $\sigma_{\rm lim}^{\rm sig}$(ab) & $\sigma_{\rm lim}^{\rm sig}$ (ab)\\
  $(\text{TeV})$   &   $(\text{TeV})$ &efficiency&&  & $ {\cal L}=3\;\text{ab}^{-1}$ & $ {\cal L}=140\;\text{fb}^{-1}$ \\
    \hline\hline
    1.5 & 1.0 & 0.0502 & 20.46 & 30.71 & 173.01 & 1018\\ \hline
    1.5 & 1.3 & 0.0572 & 37.26 & 42.09 & 186.48 & 1050\\ \hline\hline
    2.0 & 1.0 & 0.0426 & 7.21 & 7.67 & 116.59 & 817\\ \hline
    2.0 & 1.5 & 0.0581 & 9.07 & 9.07 & 93.19 & 632\\ \hline
    2.0 & 1.8 & 0.0424 & 4.08 & 5.19 & 95.52 & 731\\ \hline\hline
    2.5 & 1.0 & 0.046 & 3.83 & 4.65 & 84.95 & 662\\ \hline
    2.5 & 1.5 & 0.0482 & 2.33 & 2.33 & 63.6 & 564\\ \hline
    2.5 & 1.8 & 0.0493 & 2.28 & 2.1 & 60.64 & 545\\ \hline
    2.5 & 2.3 & 0.0328 & 0.47 & 0.74 & 57.95 & 709\\ \hline
  \end{tabular}
  \end{center}
  \caption{$pp \to 2 \ell 2 b$: signal efficiency, background cross sections, and expected $pp\to H \to b_4 \bar b \to Z b\bar b \to \ell^+\ell^-b\bar b$ cross section upper limits for various mass configurations.}
  \label{tab:ZbbVals}
\end{table}
\begin{figure}[t]
  \begin{center}
    \includegraphics[width=0.4 \linewidth]{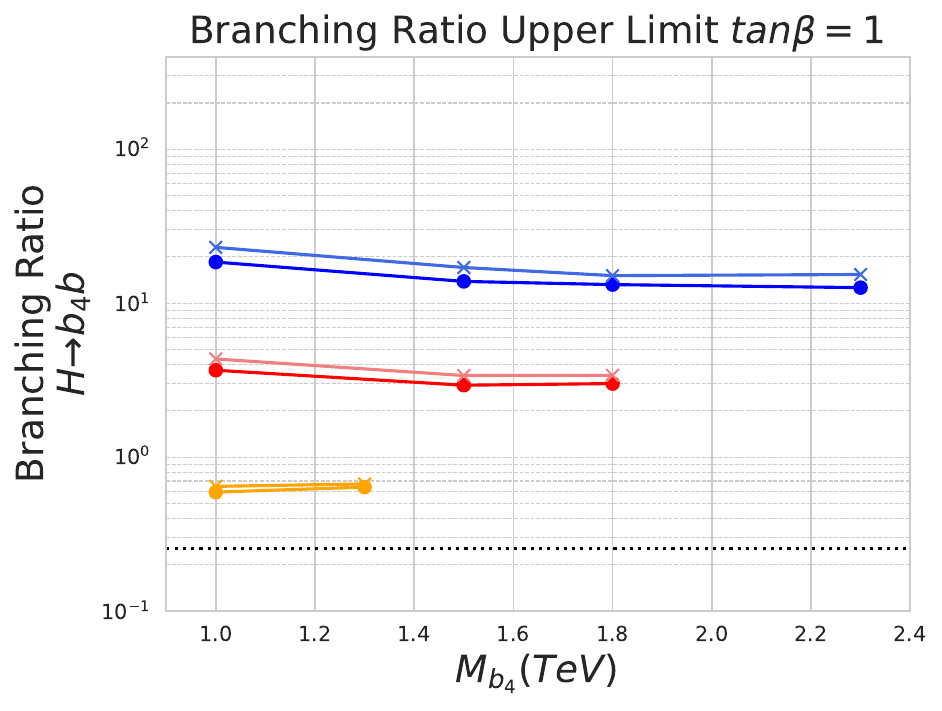}
    \includegraphics[width=0.4 \linewidth]{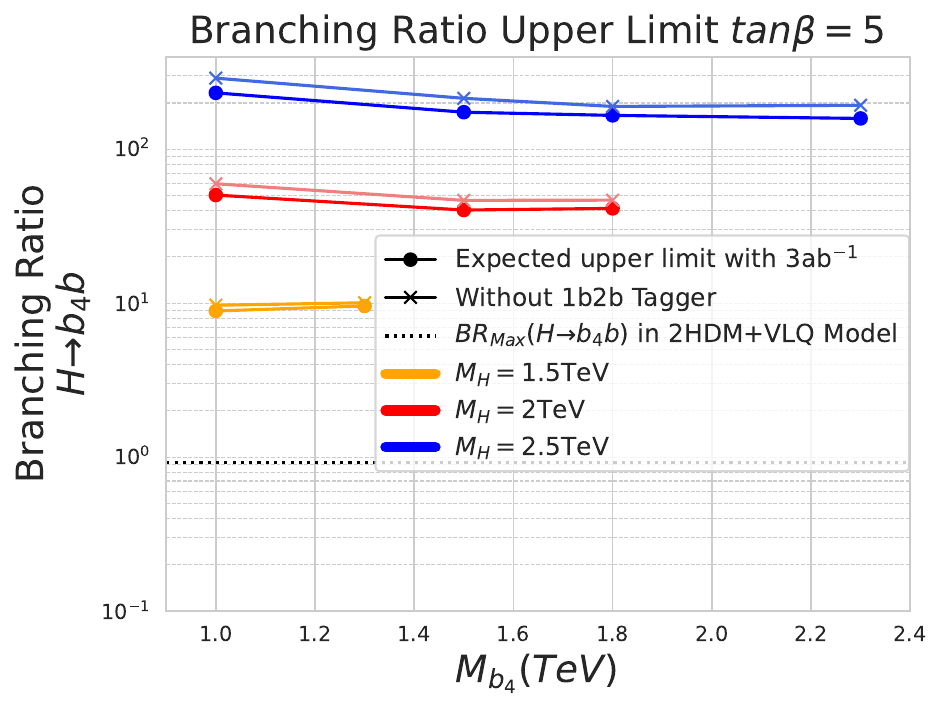}
    \includegraphics[width=0.4 \linewidth]{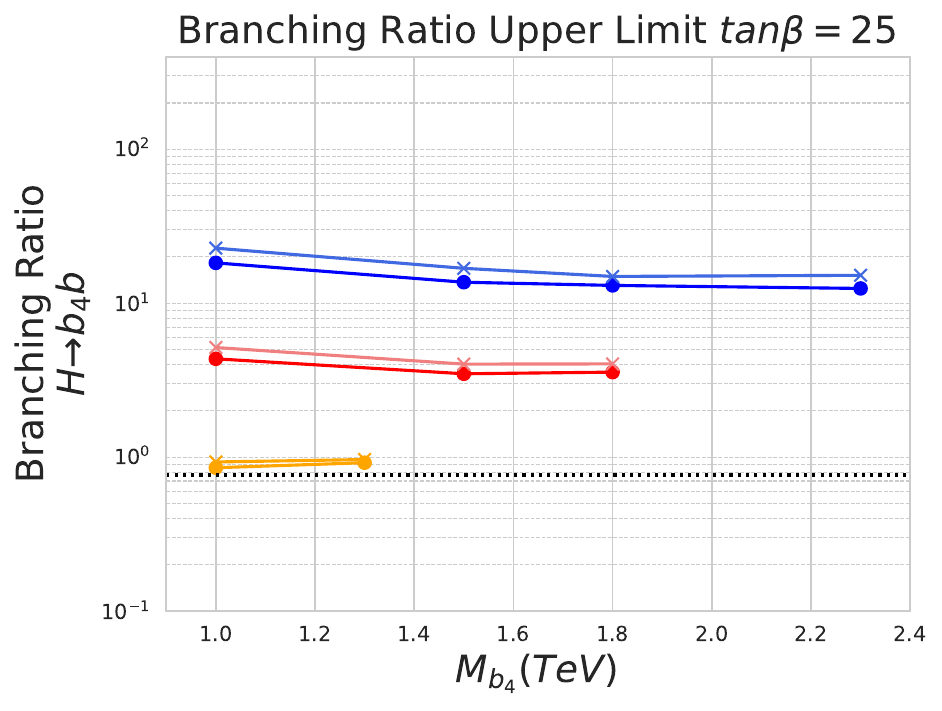}
    \includegraphics[width=0.4 \linewidth]{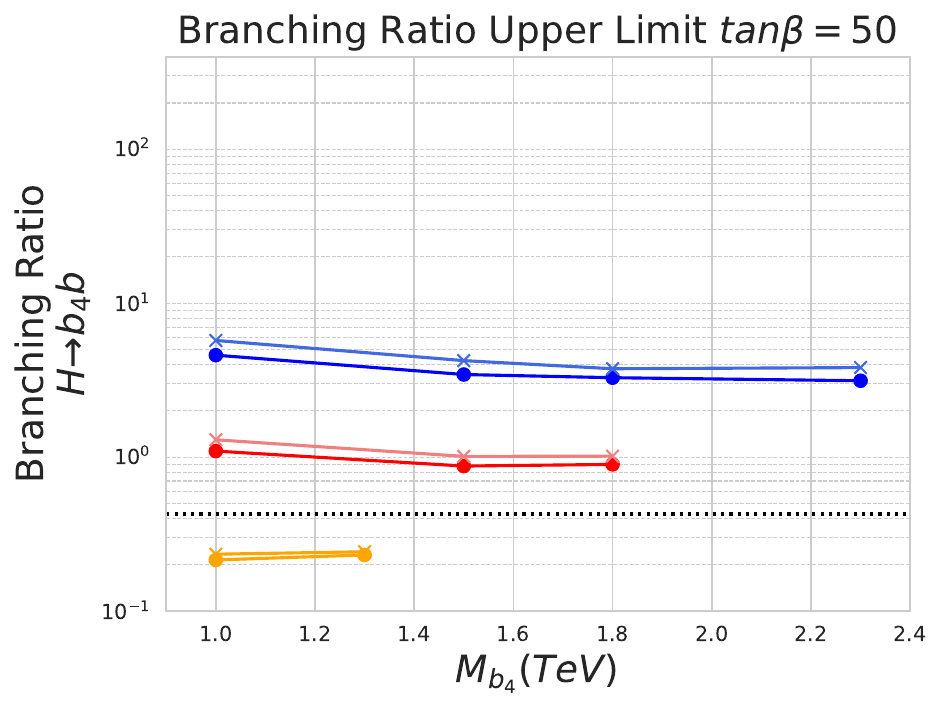}
  \end{center}
  \caption{$pp\to 2\ell 2b$: upper bounds on Branching Ratio on the branching ratio $H\to b_4 b$ for $\tan\beta=1,5,25,50$.}
  \label{fig:BRZbb}
\end{figure}

The optimal cuts that we find for the various mass configurations we consider are summarized in table~\ref{tab:ZbbCuts} and the corresponding expected upper limits on the $pp \to H \to b_4 \bar b$ cross section are presented in table~\ref{tab:ZbbVals} and figure~\ref{fig:CSupZbb}. Note that we take ${\rm BR} (b_4 \to Z b) \simeq 50\%$ in accordance to the Goldstone boson equivalence limit (see, for instance, Table~3 of ref.~\cite{Dermisek:2019vkc}). In table~\ref{tab:ZbbVals} we present expected upper limits for total integrated luminosities of 140 fb${}^{-1}$ and 3 ab${}^{-1}$.

As it appears from the $M_{\ell\ell j_0}$ distributions presented in figures~\ref{fig:MZbb_1510}--\ref{fig:MZbb_2523}, the main impact of the 1b2b-tagger is to reduce considerably the irreducible background from $pp \to jj \ell^+\ell^-$.

In the context of the model we are considering, the signal cross section upper limits can be converted into constraints on the $H\to b_4 b \to Zb\bar b $ branching ratio at fixed values of $\tan\beta$. The explicit form of the maximum possible cross section for $p p \to H \to b_4 b$ in the models we consider correspond to the solid lines in figure~\ref{fig:CSupZbb_Theory} which have been generated using the data used in ref.~\cite{Dermisek:2019heo} (in figure~4 of ref.~\cite{Dermisek:2019heo} the $M_H = 2.5$ TeV case is explicitly presented). The combination of the maximum theoretical cross section with the explicit upper limits presented in figure~\ref{fig:CSupZbb} (which are indicated as horizontal lines in figure~\ref{fig:CSupZbb_Theory}) yields the expected upper limits on the branching ratio $H\to b_4 b$ summarized in figure~\ref{fig:BRZbb}.

\section{Conclusion}
\label{sec:Conclusion}
We presented a detailed analysis of the expected sensitivity of the LHC experiments to new physics signals involving vectorlike quarks and heavy Higgses which couple to the third generation of SM quarks. When both type of particles are simultaneously present, they dominantly decay into each other (depending on their mass hierarchy) leading to signatures with large multiplicities of $b$-jets. We focused on the decay chains
\begin{alignat}{4}
pp &\to b_4 \bar b_4      \to (Hb) (H \bar b) &&\to (b\bar b b) (b\bar b \bar b) \; , \\ 
pp &\to H \to b_4 \bar b \to  Z b  \bar b  &&\to \ell^+ \ell^- b \bar b \;\;\; (\ell = e,\mu) \; ,
\end{alignat}
and presented a detailed study of the LHC experiments with current ($140 \text{fb}^{-1}$) and projected ($3 \text{ab}^{-1}$) integrated luminosities. An important tool in the analysis is the implementation of a tagger which aims at differentiating $b$-tagged jets which originated from a $b$-quark vs a gluon (1b2b-tagger).  We found sensitivity to vectorlike quarks masses up to about 2.1 TeV in the $6b$ final state (see, figure~\ref{fig:BRup6b}) and to heavy Higgs masses up to 1.5 TeV in the $2b2\ell$ one (see figure~\ref{fig:BRZbb}). In order to reach these sensitivities, the 1b2b tagger would have to be explicitly implemented by ATLAS and CMS. Finally, we stress that a complete simulation of the multi-jet background originated from light flavors (4j) would require a number of simulated events which beyond our current computing capabilities ($b$-tagging alone costs about 7\% per light-flavor initated jet) and that data driven techniques most likely will be needed.

\begin{figure}[h]
  \begin{center}
    \includegraphics[width=0.75 \linewidth]{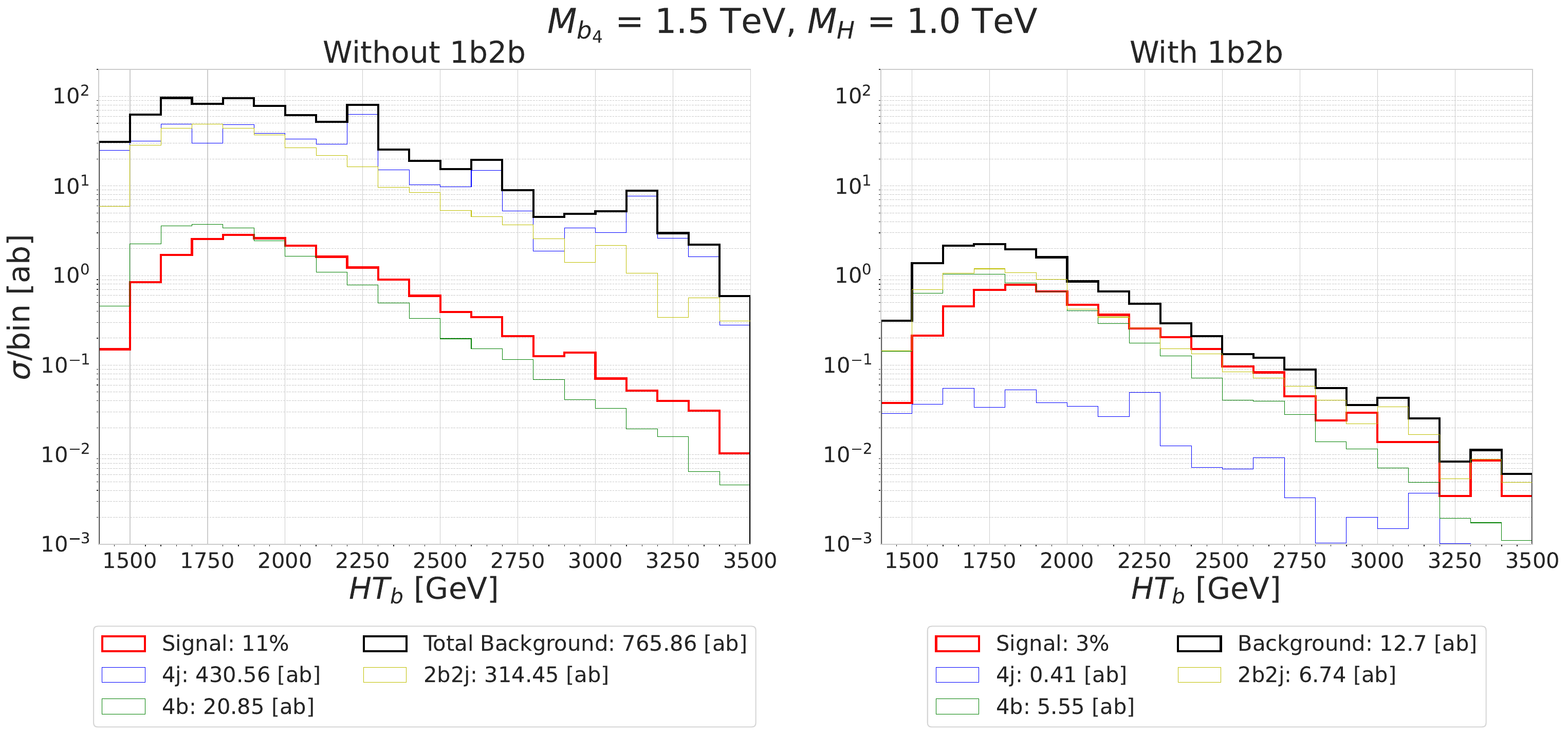}
    \includegraphics[width=0.75 \linewidth]{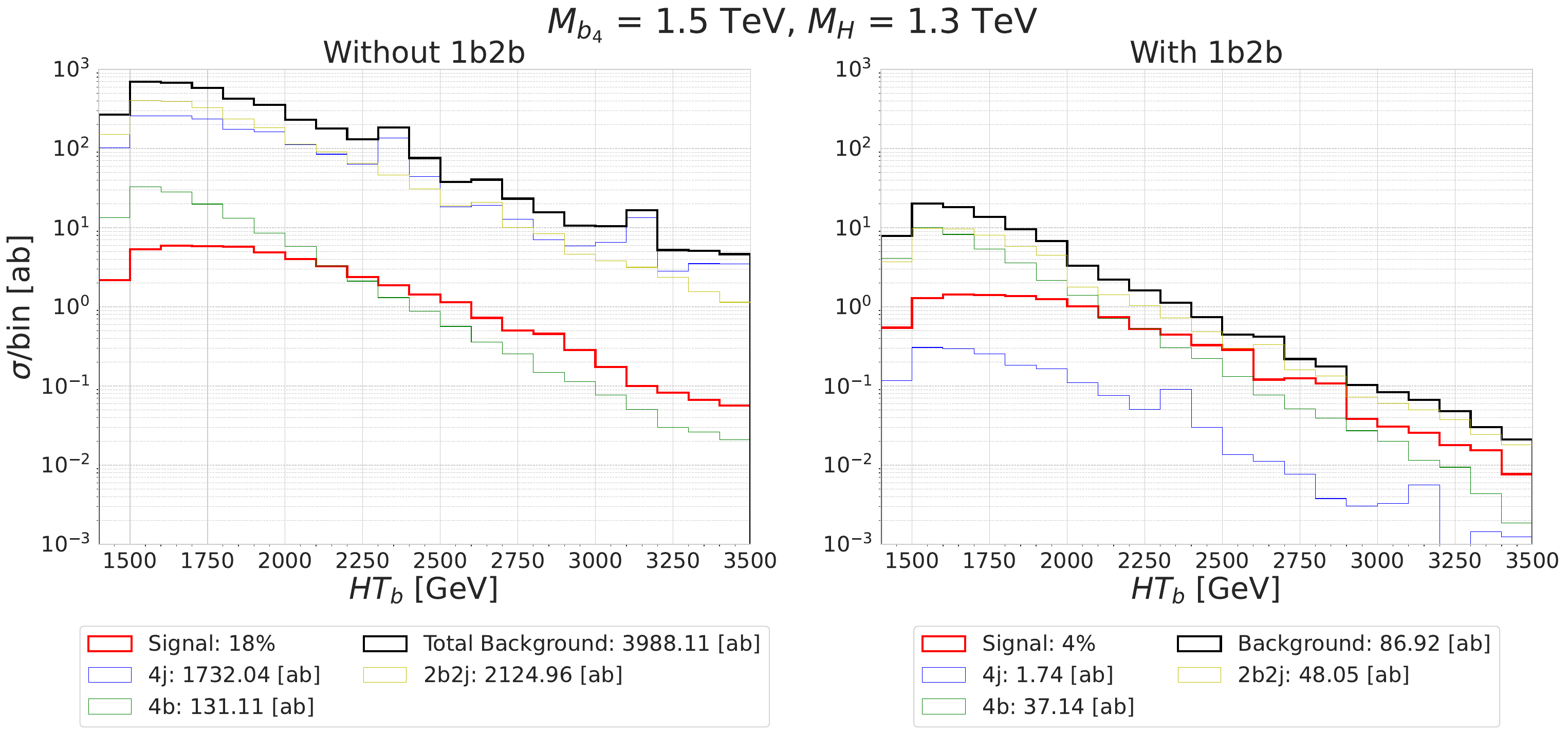}
      \includegraphics[width=0.75 \linewidth]{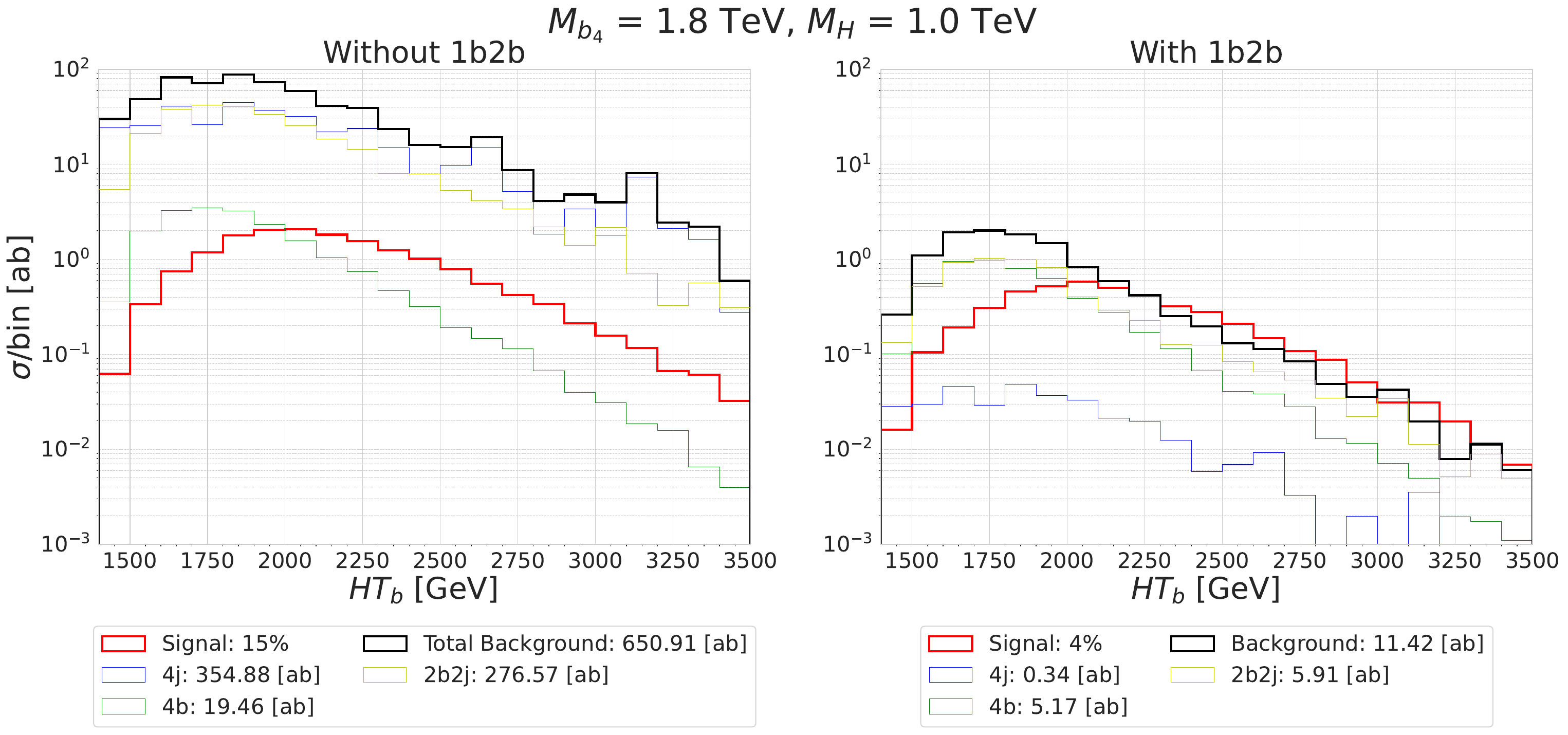}
  \includegraphics[width=0.75 \linewidth]{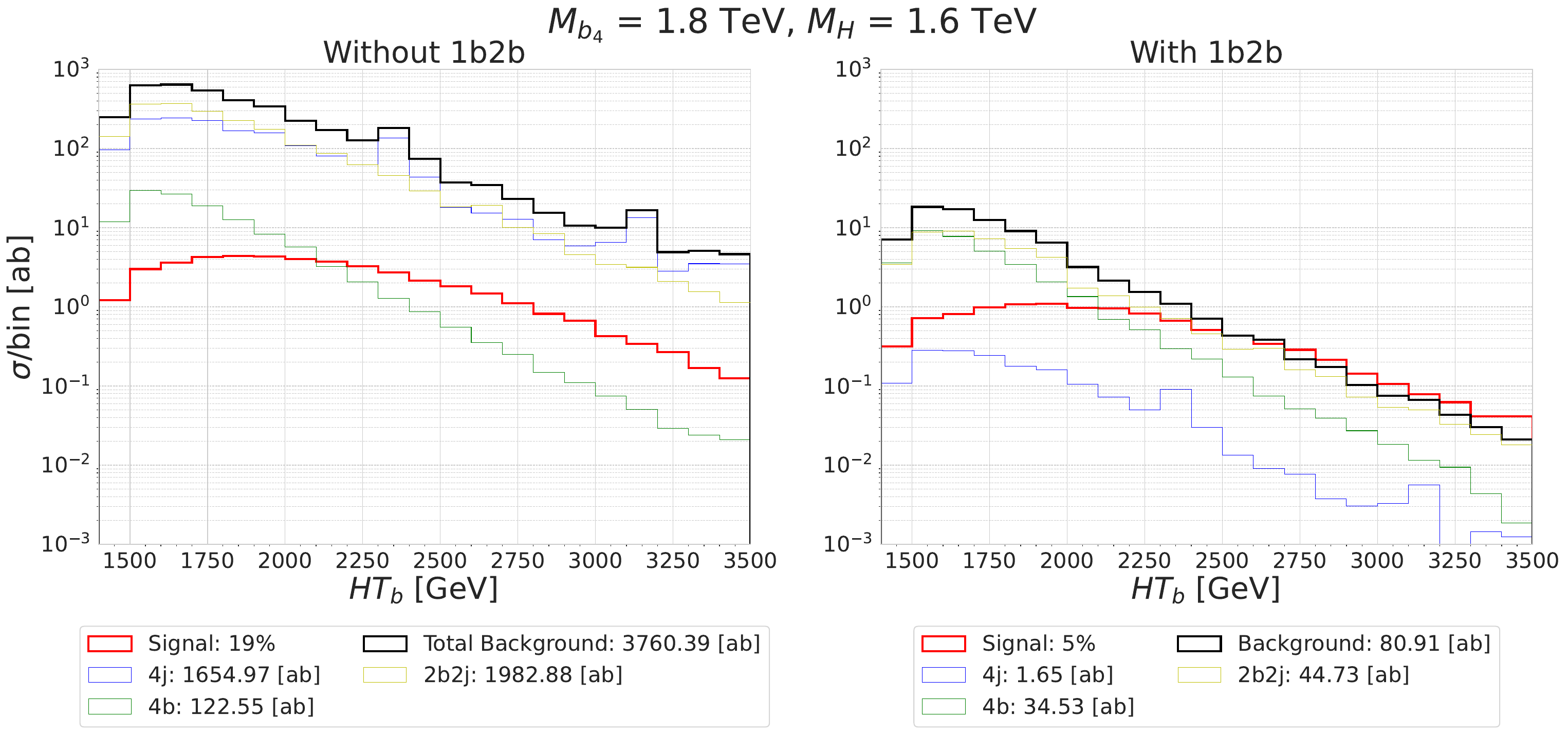}
  \end{center}
\caption{$pp\to 6b$: $H_{T_b}$ distributions for various $b_4$ and $H$ mass configurations after all cuts from table~\ref{tab:6bCuts} have been applied. Left panels show the results prior to the application of the 1b2b tagger.}
\label{fig:HTb1}
\end{figure}

\begin{figure}[h]
  \begin{center}
  \includegraphics[width=0.75 \linewidth]{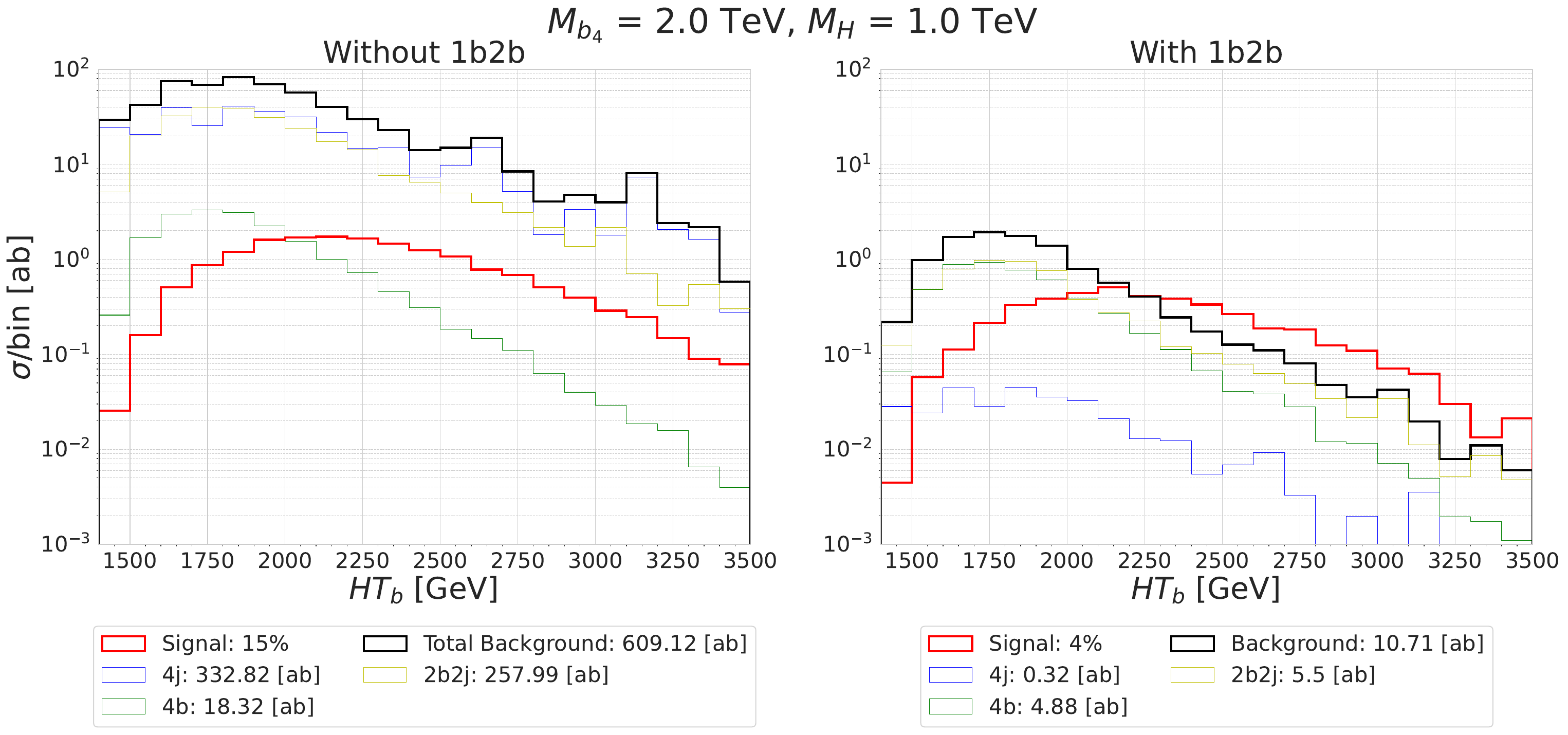}
  \includegraphics[width=0.75 \linewidth]{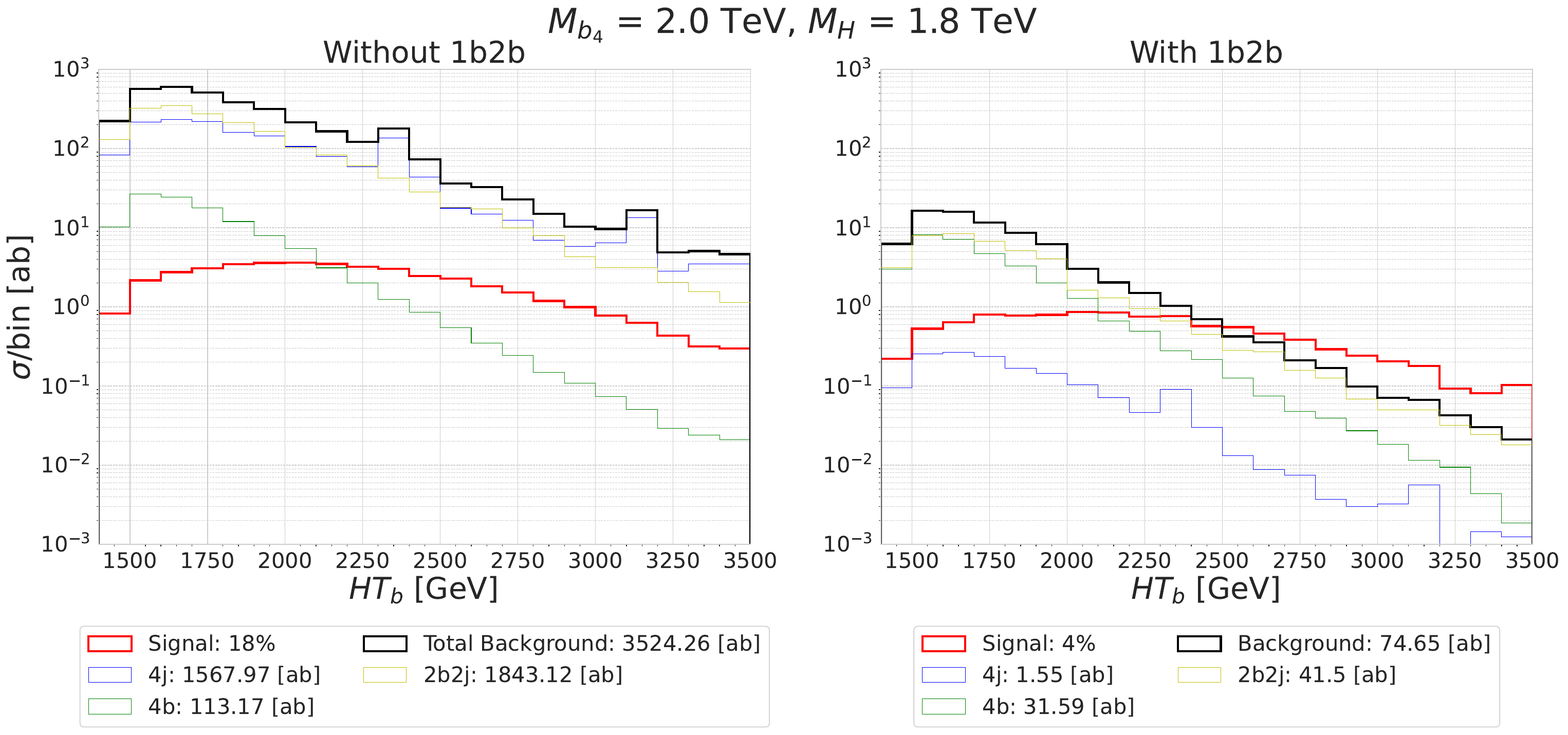}
  \includegraphics[width=0.75 \linewidth]{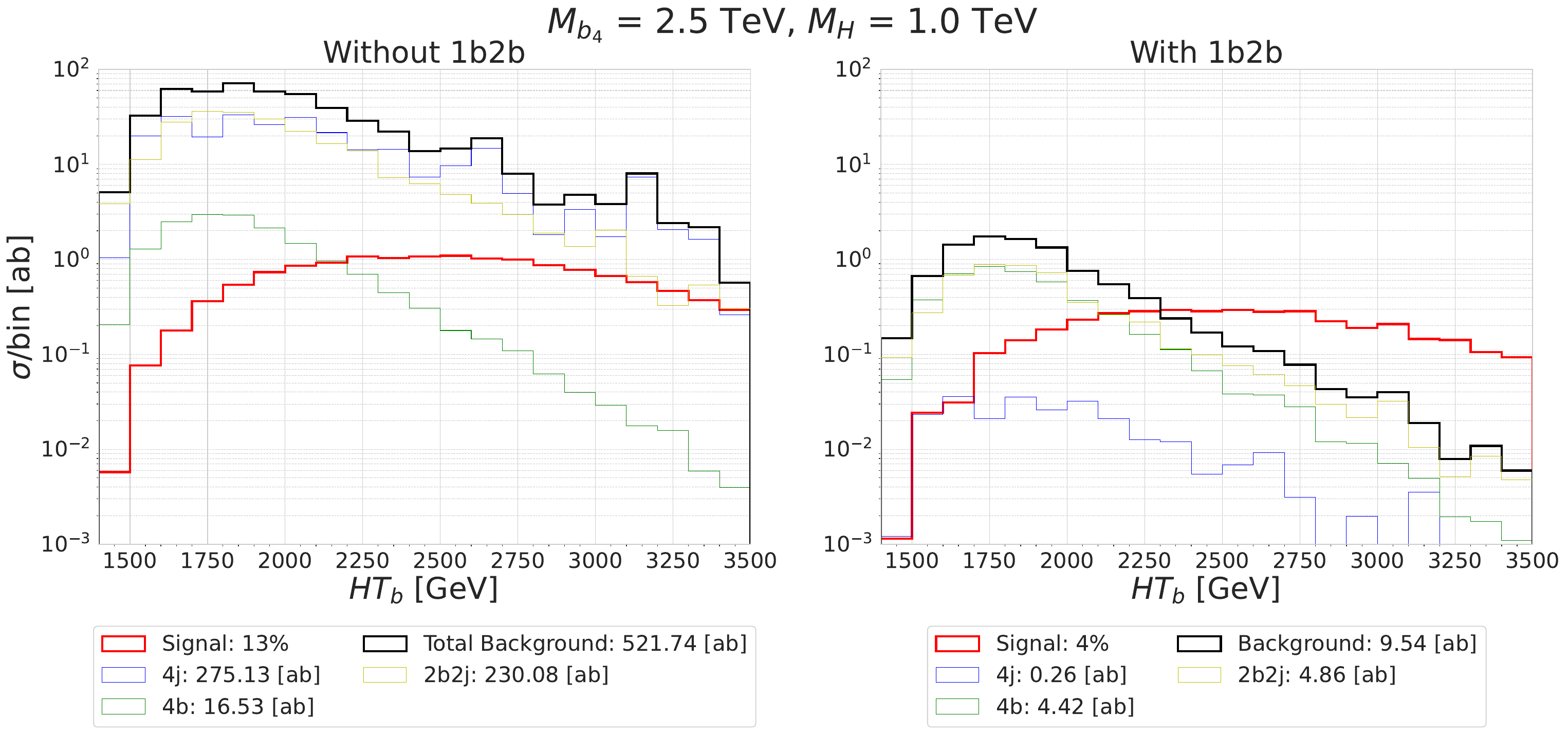}
  \includegraphics[width=0.75 \linewidth]{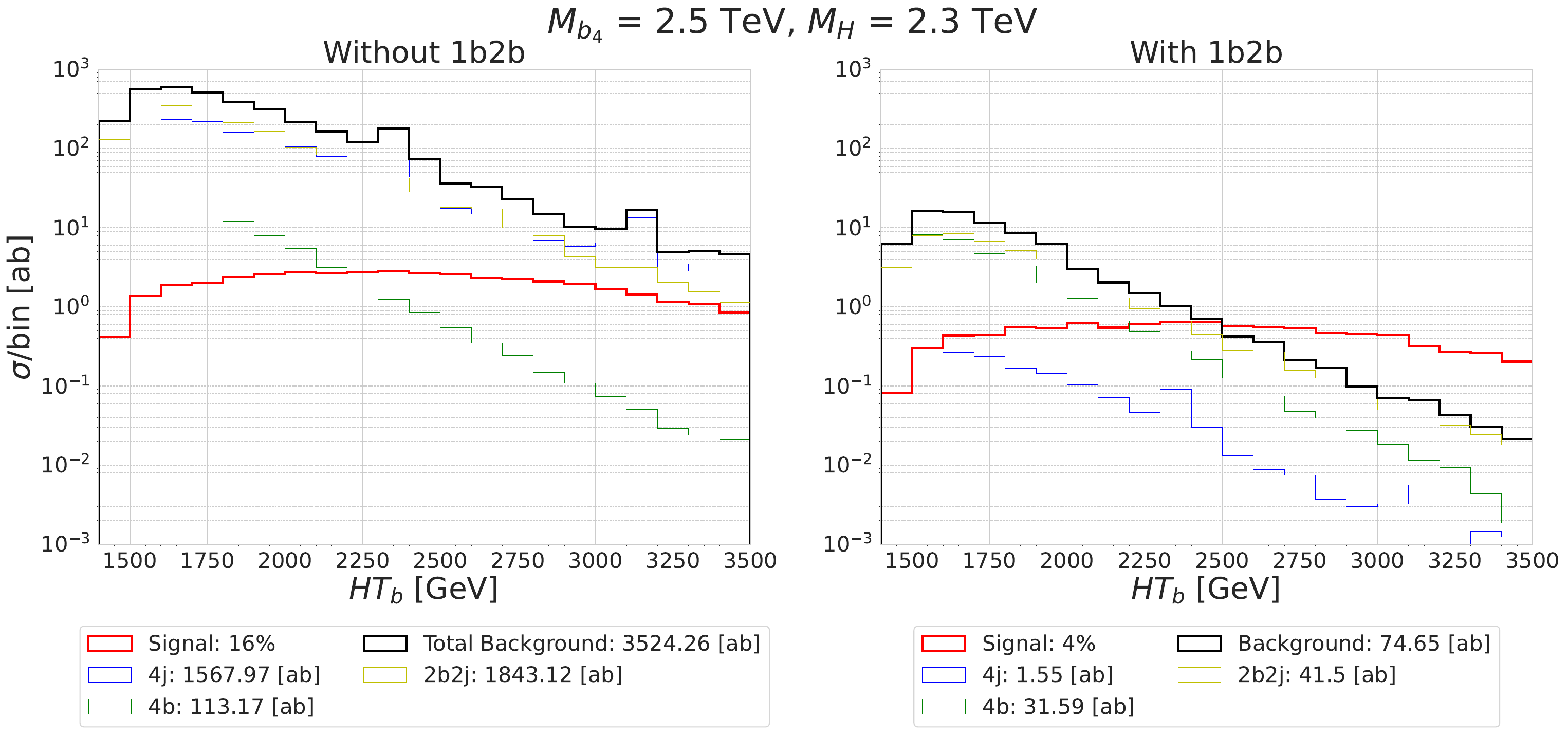}
  \end{center}
\caption{$pp\to 6b$: $H_{T_b}$ distributions for various $b_4$ and $H$ mass configurations after all cuts from table~\ref{tab:6bCuts} have been applied. Left panels show the results prior to the application of the 1b2b tagger.}
\label{fig:HTb2}
\end{figure}

\begin{figure}[h]
\begin{center}
  \includegraphics[width=0.75 \linewidth]{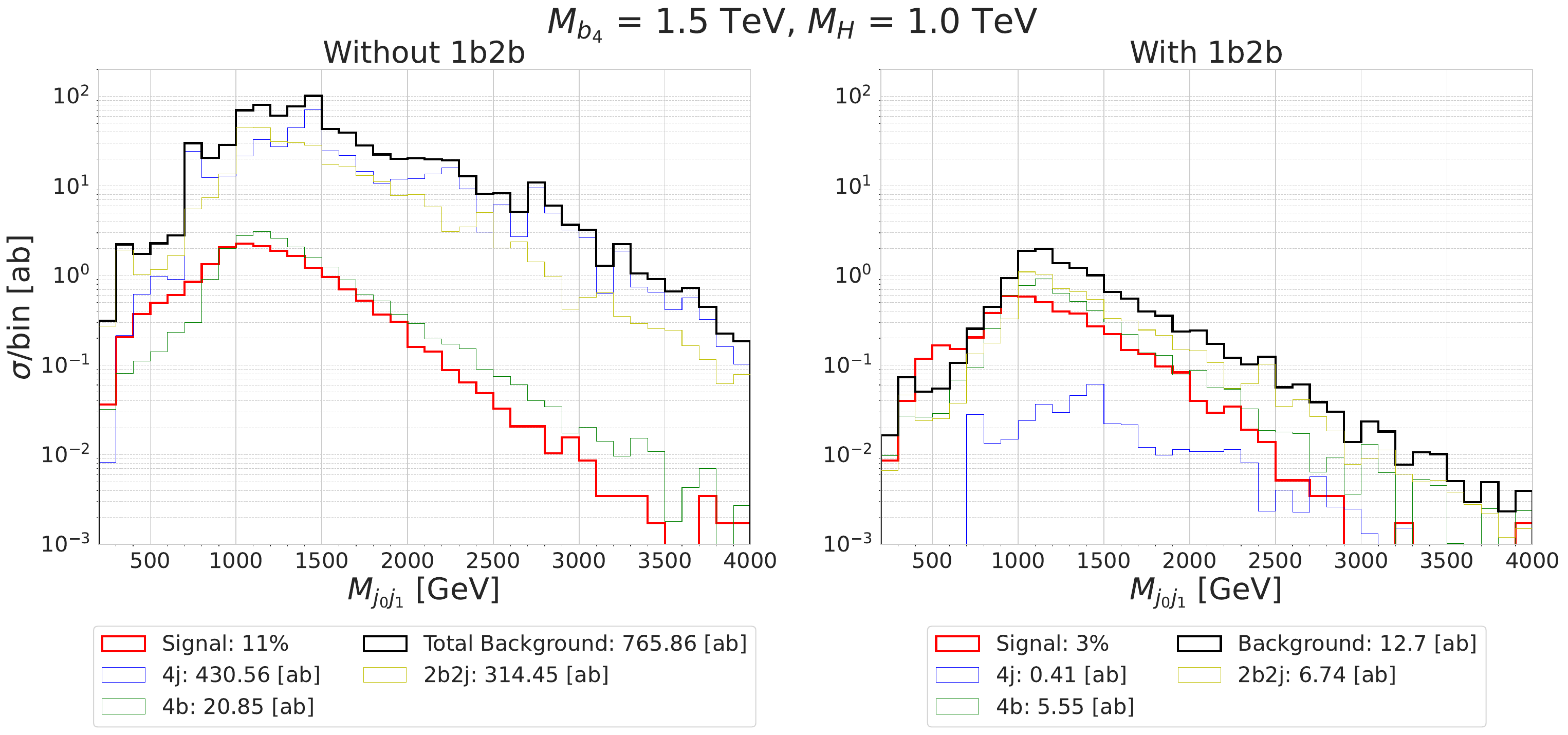}
  \includegraphics[width=0.75 \linewidth]{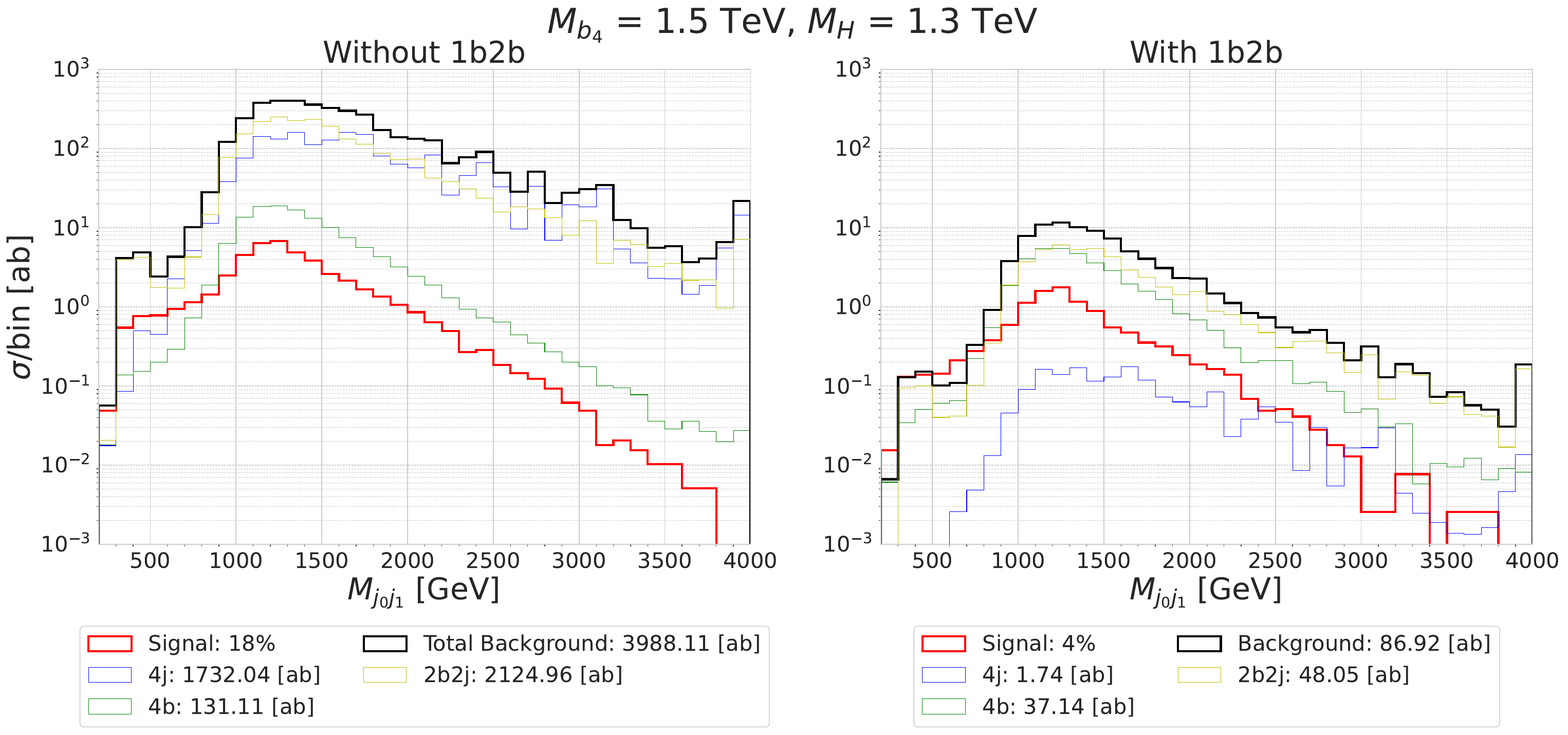}
  \includegraphics[width=0.75 \linewidth]{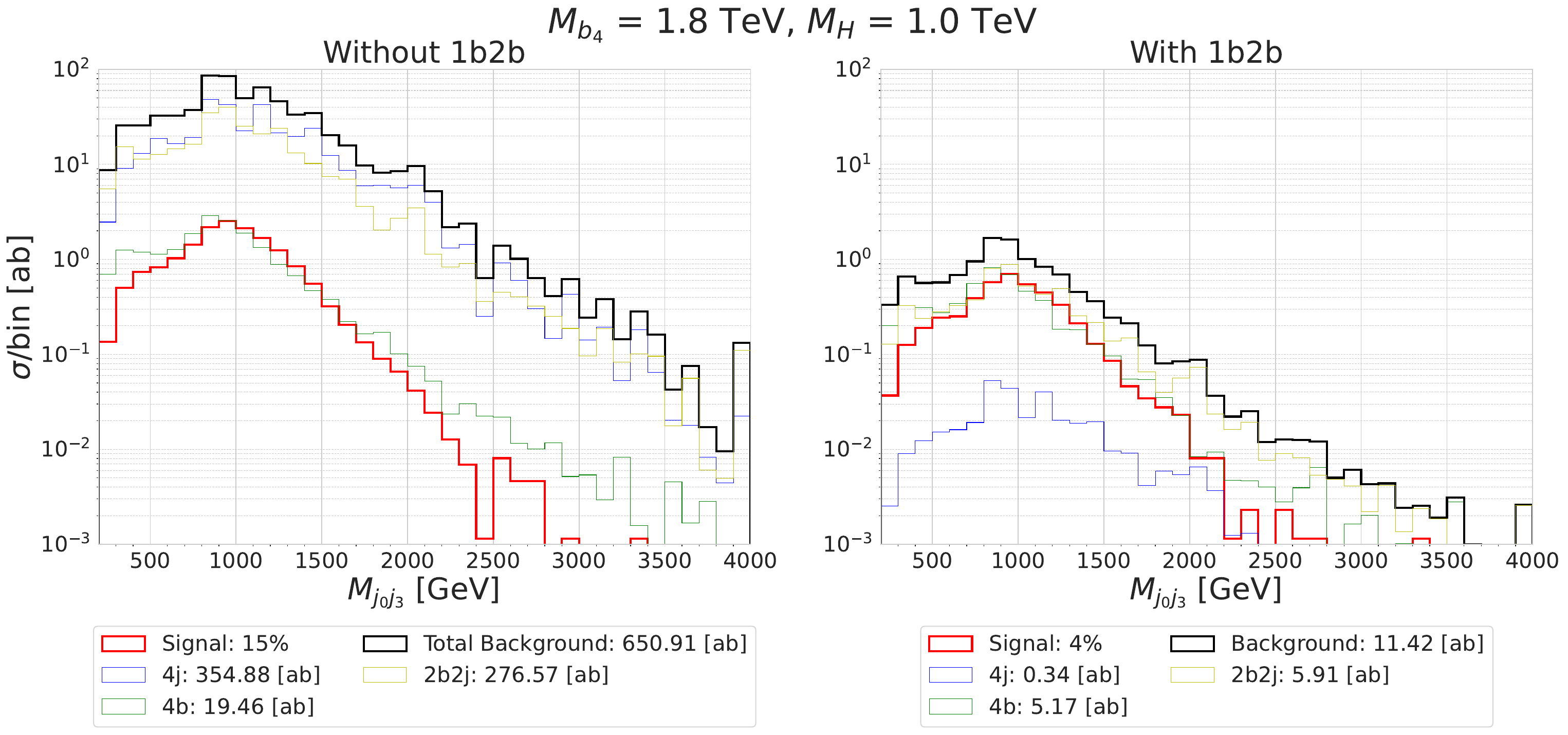}
  \includegraphics[width=0.75 \linewidth]{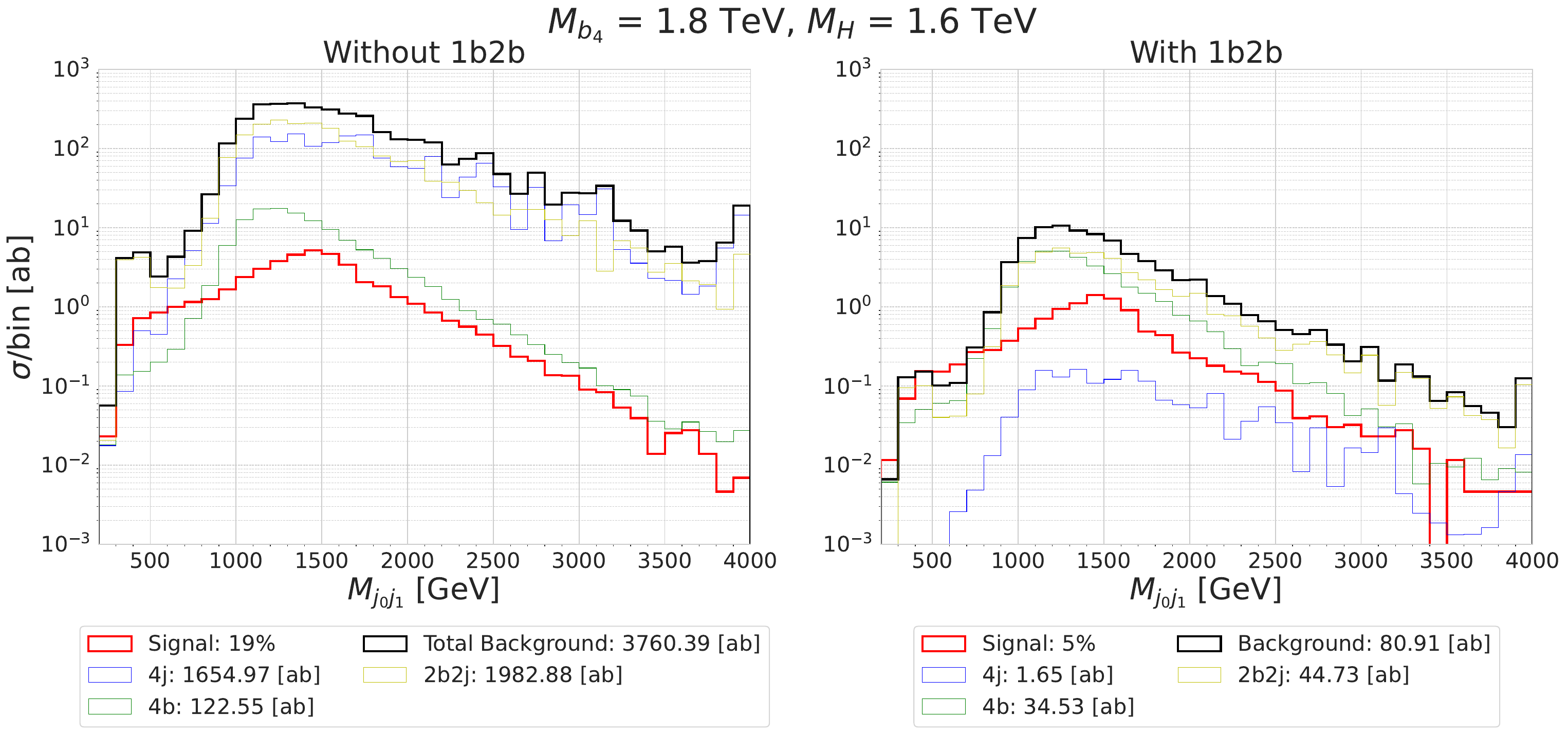}
\end{center}
\caption{$pp\to 6b$: di-jet and tri-jet invariant mass distributions for various $b_4$ and $H$ mass configurations after all cuts from table~\ref{tab:6bCuts} have been applied. Left panels show the results prior to the application of the 1b2b tagger.}
\label{fig:MJJ1}
 \end{figure}

\begin{figure}[h]
\begin{center}
  \includegraphics[width=0.75 \linewidth]{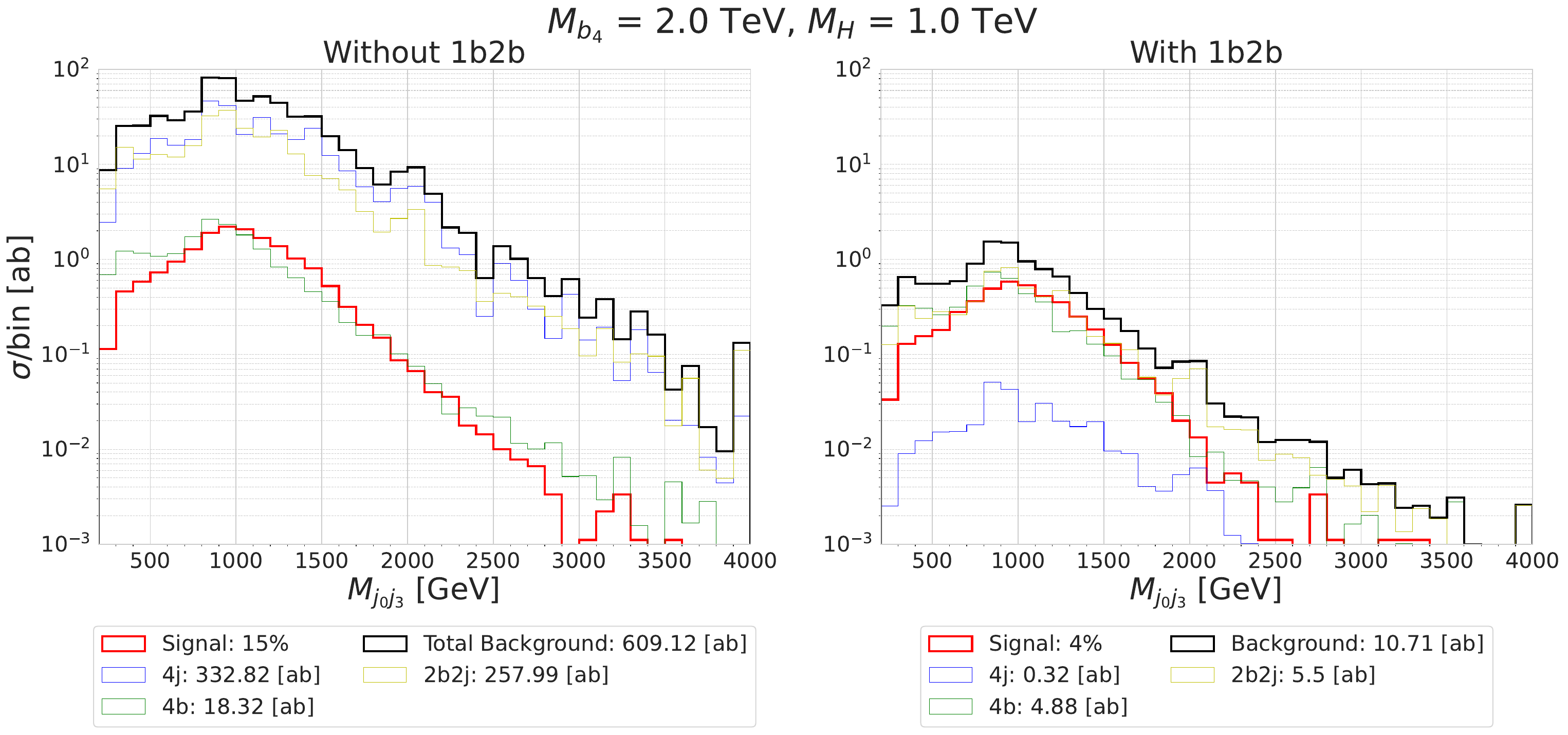}
  \includegraphics[width=0.75 \linewidth]{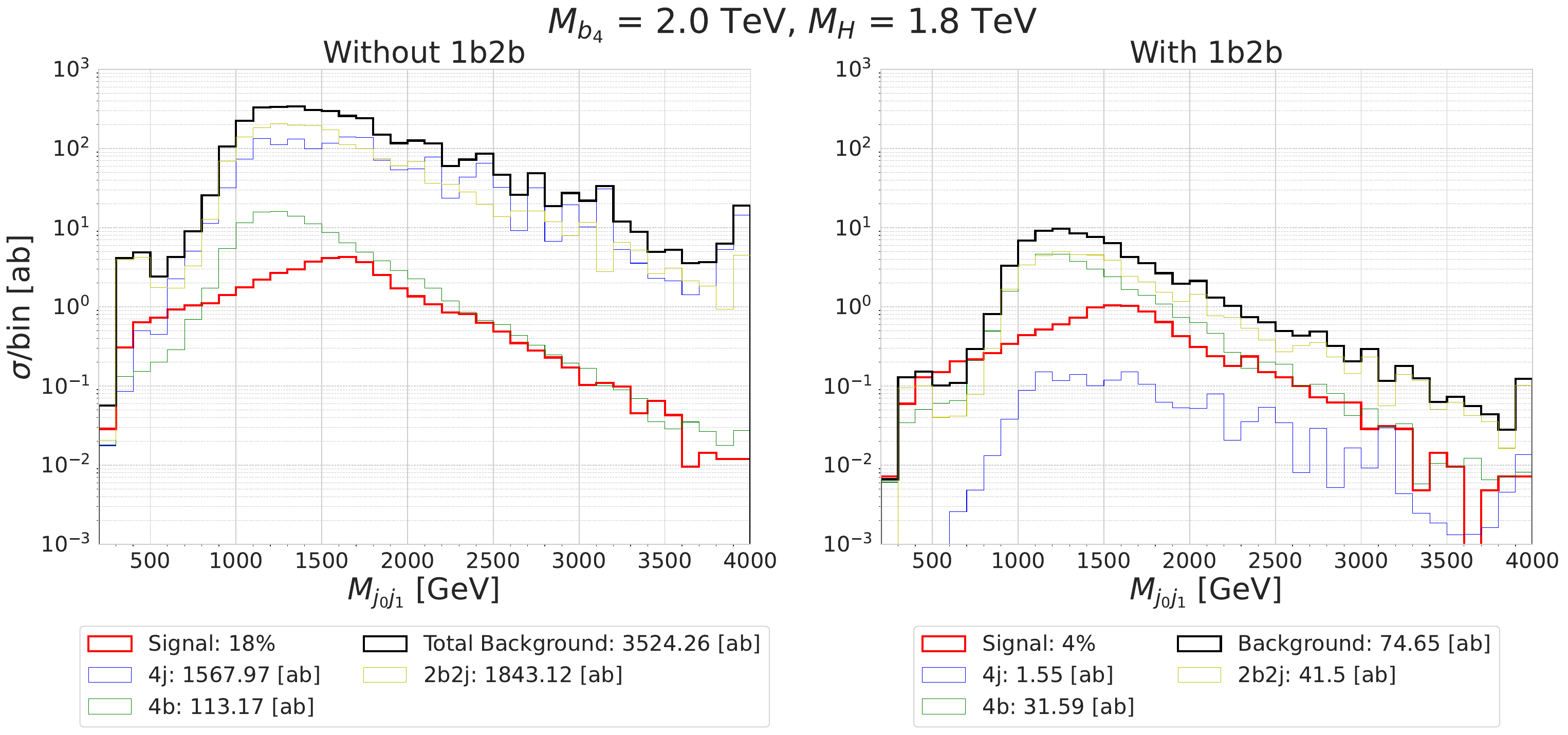}
  \includegraphics[width=0.75 \linewidth]{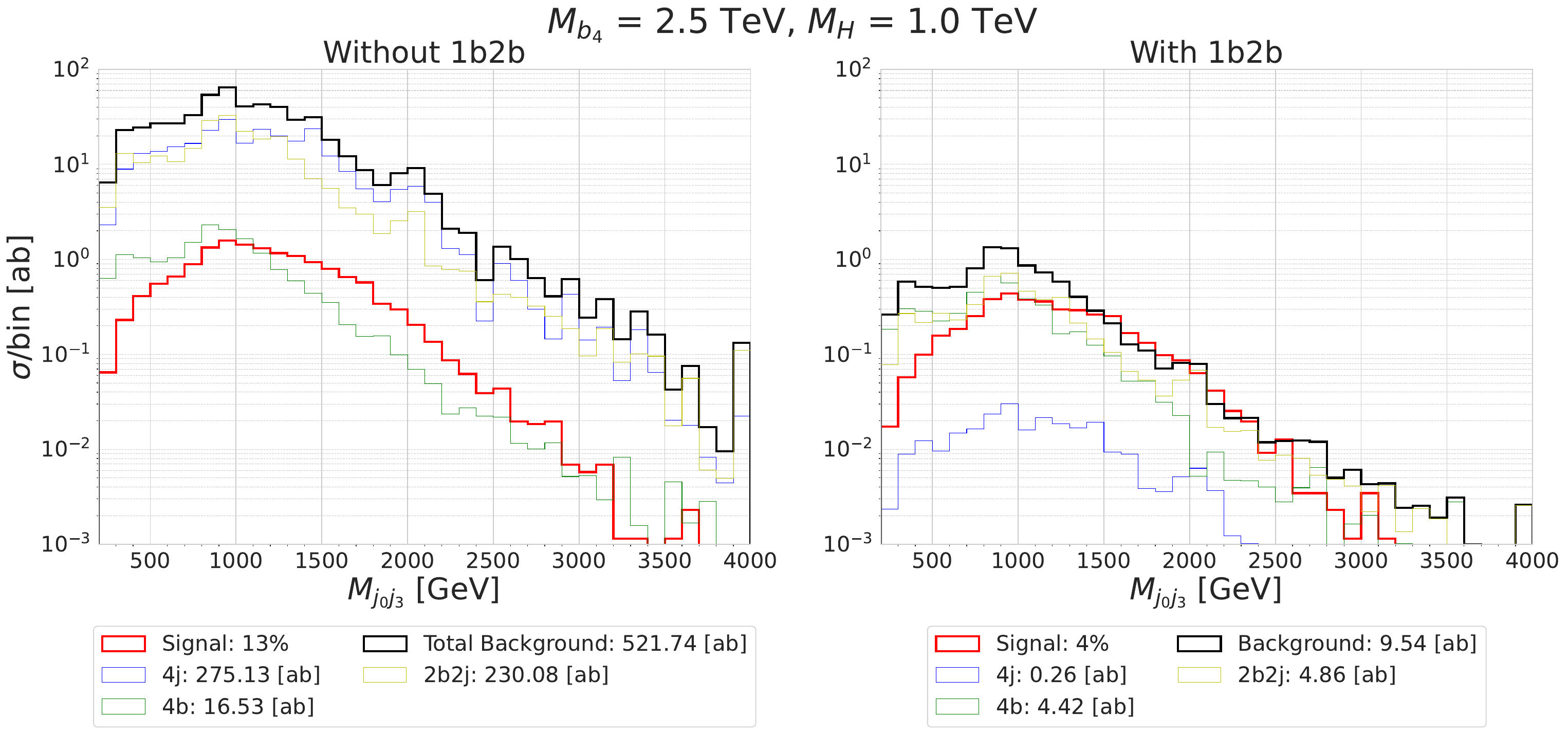}
  \includegraphics[width=0.75 \linewidth]{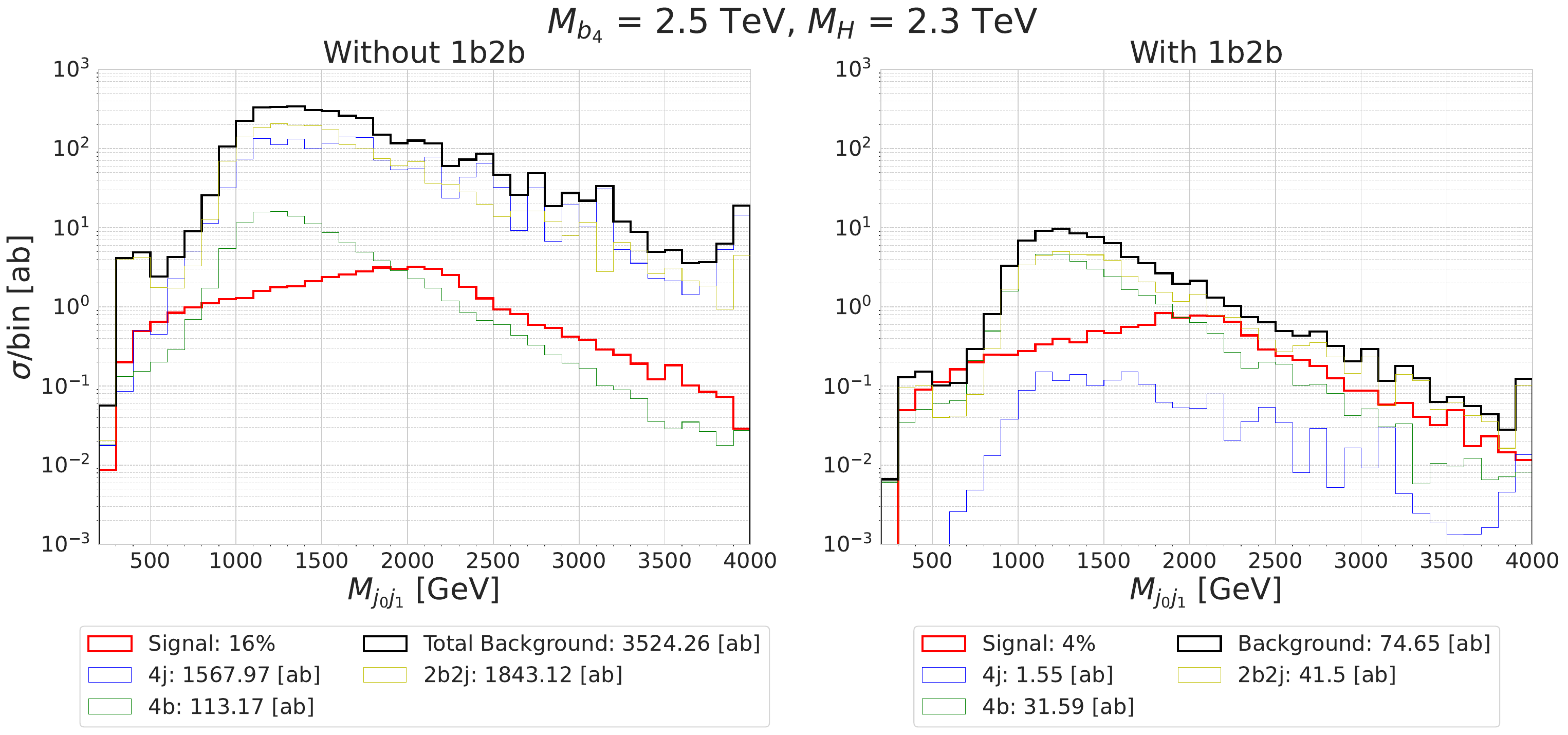}
\end{center}
\caption{$pp\to 6b$: di-jet and tri-jet invariant mass distributions for various $b_4$ and $H$ mass configurations after all cuts from table~\ref{tab:6bCuts} have been applied. Left panels show the results prior to the application of the 1b2b tagger.}
\label{fig:MJJ2}
 \end{figure}

\begin{figure}[h]
\begin{center}
 \includegraphics[width=0.75 \linewidth]{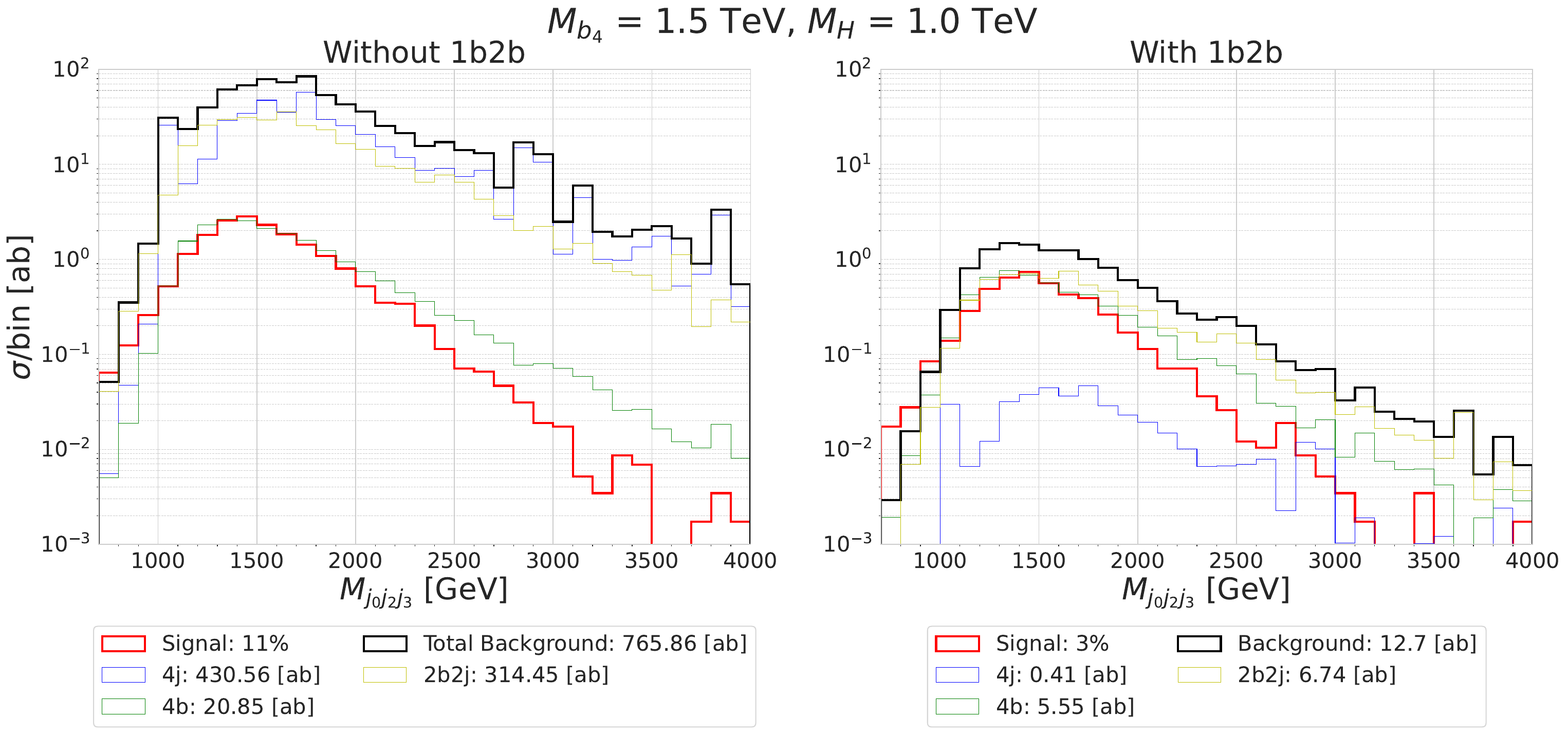}
\includegraphics[width=0.75 \linewidth]{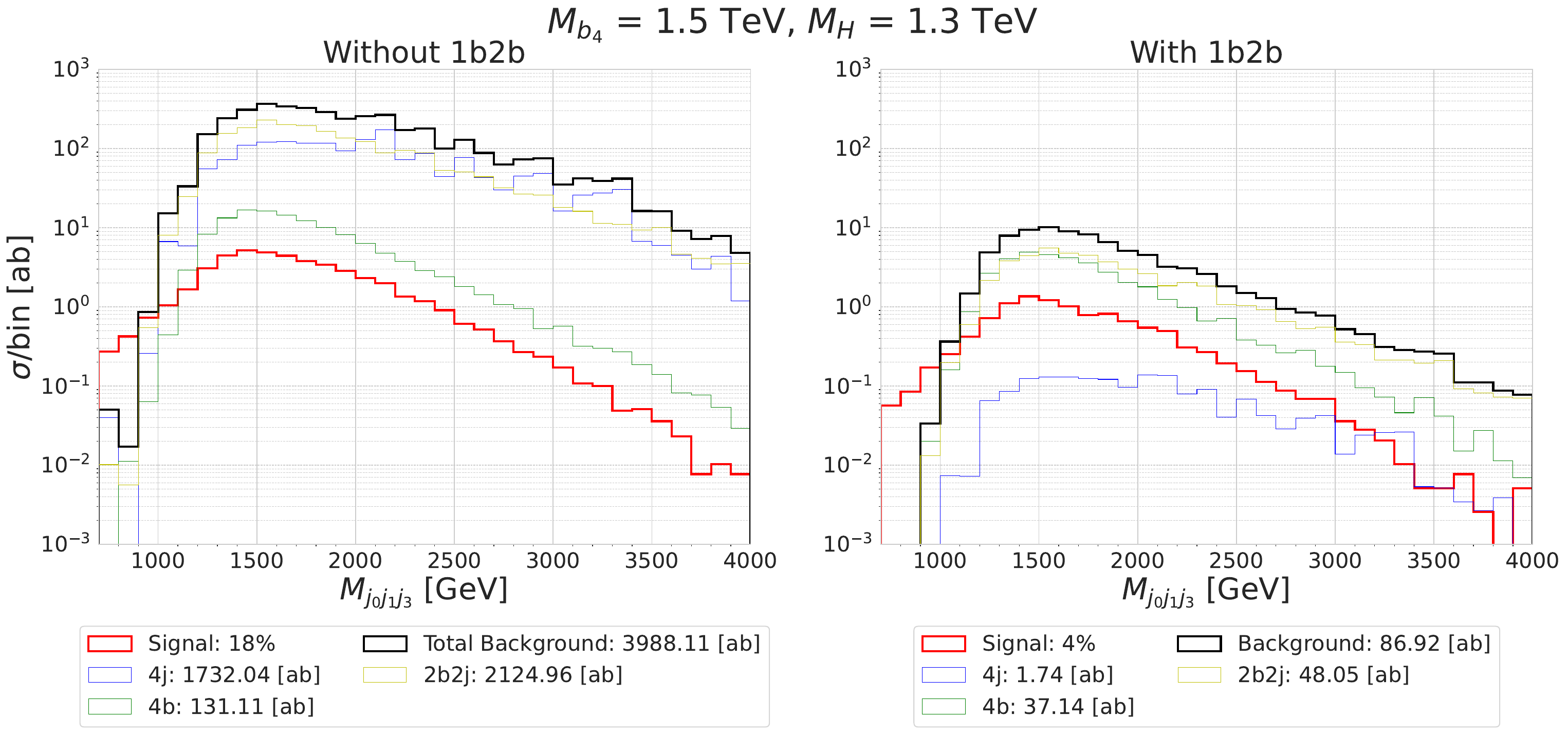}
  \includegraphics[width=0.75 \linewidth]{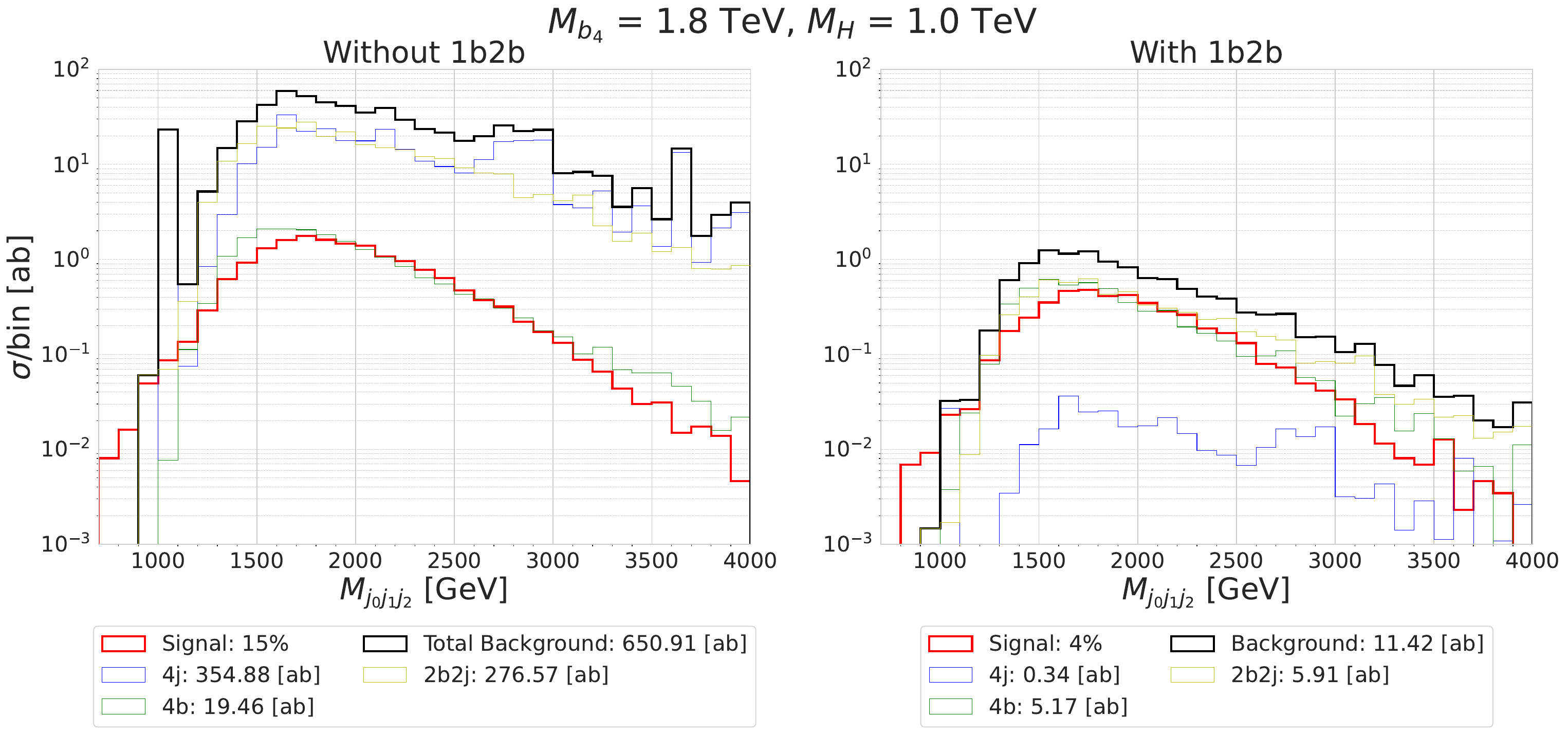}
  \includegraphics[width=0.75 \linewidth]{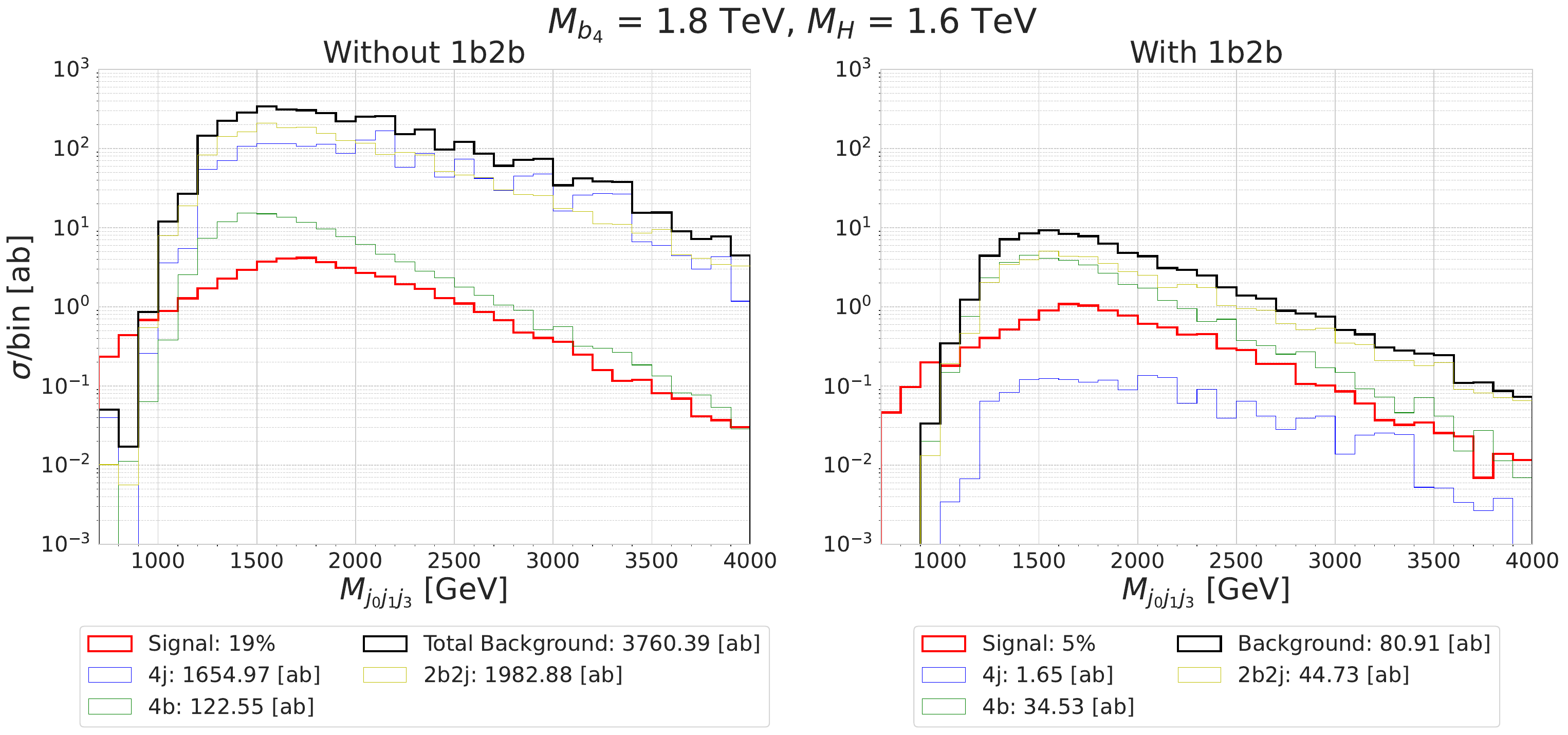}
\end{center}
\caption{$pp\to 6b$: di-jet and tri-jet invariant mass distributions for various $b_4$ and $H$ mass configurations after all cuts from table~\ref{tab:6bCuts} have been applied. Left panels show the results prior to the application of the 1b2b tagger.}
\label{fig:MJJJ1}
\end{figure}

\begin{figure}[h]
\begin{center}
  \includegraphics[width=0.75 \linewidth]{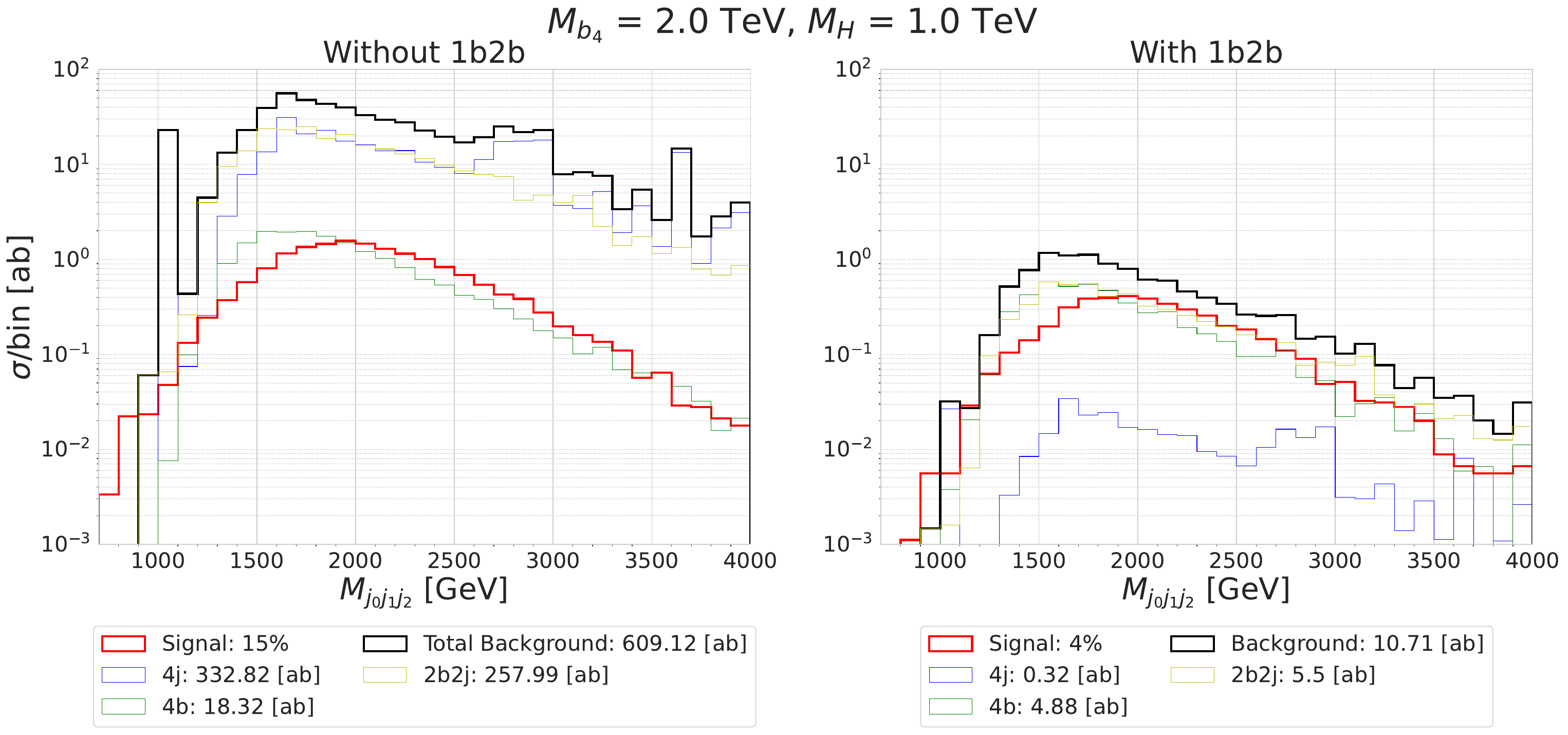}
  \includegraphics[width=0.75 \linewidth]{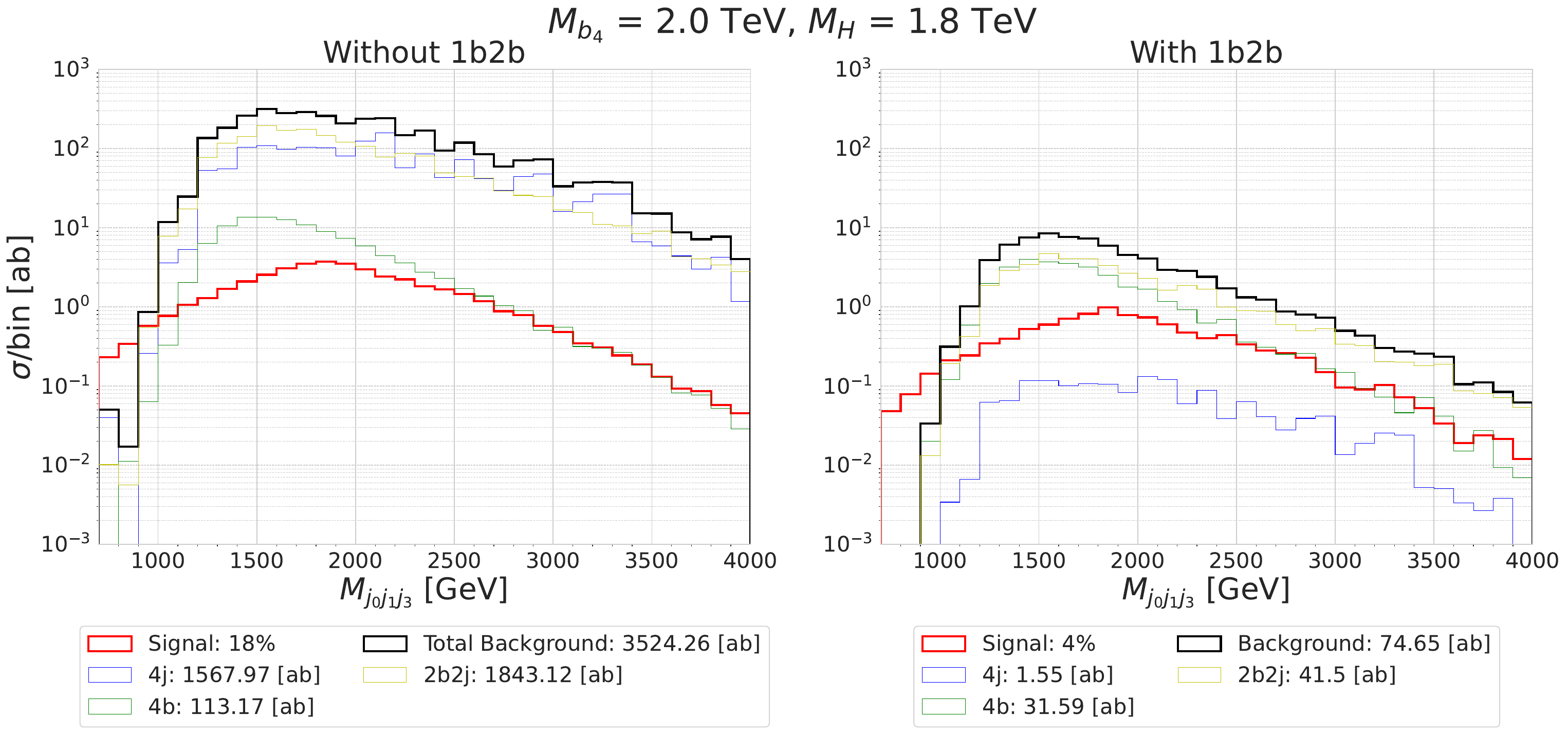}
  \includegraphics[width=0.75 \linewidth]{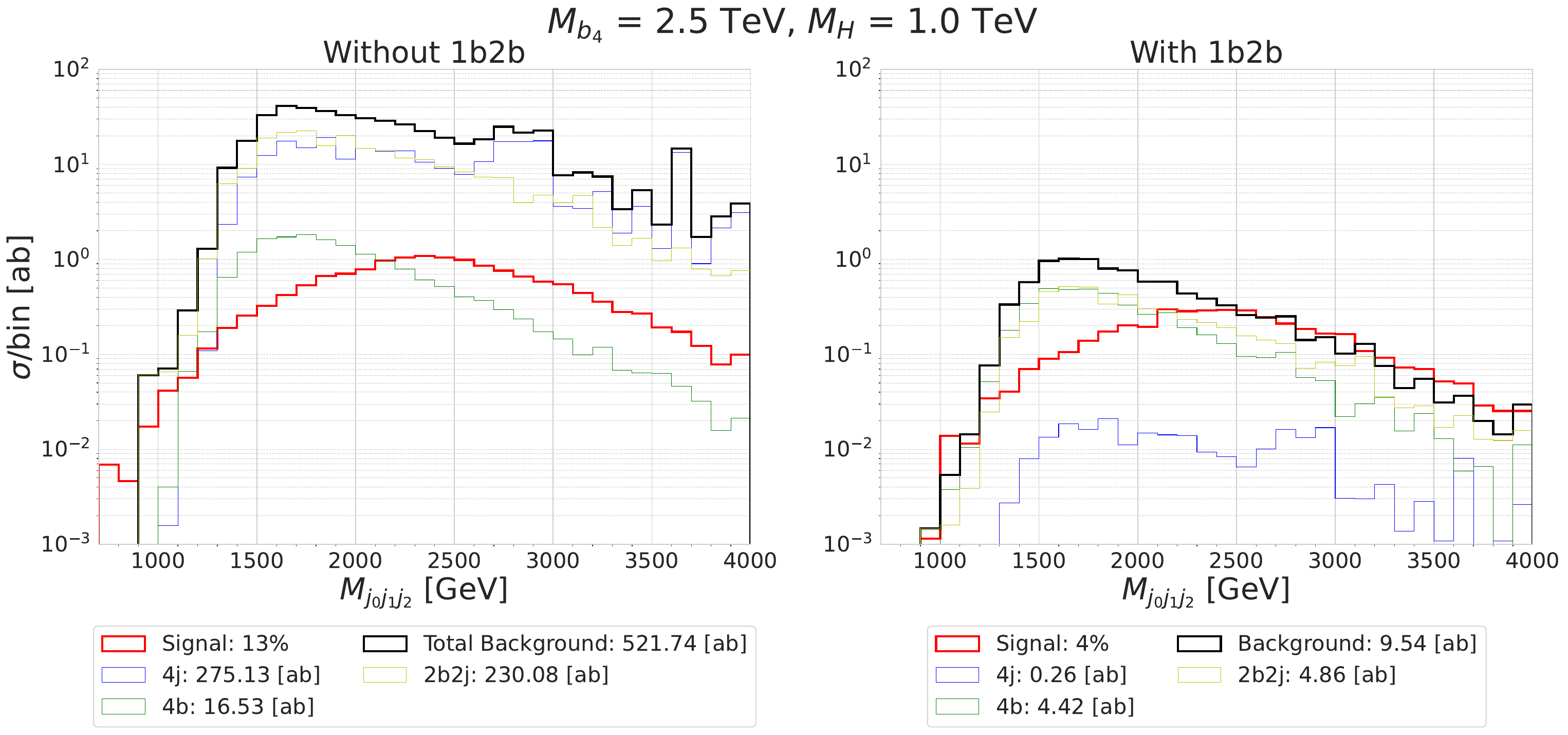}
  \includegraphics[width=0.75 \linewidth]{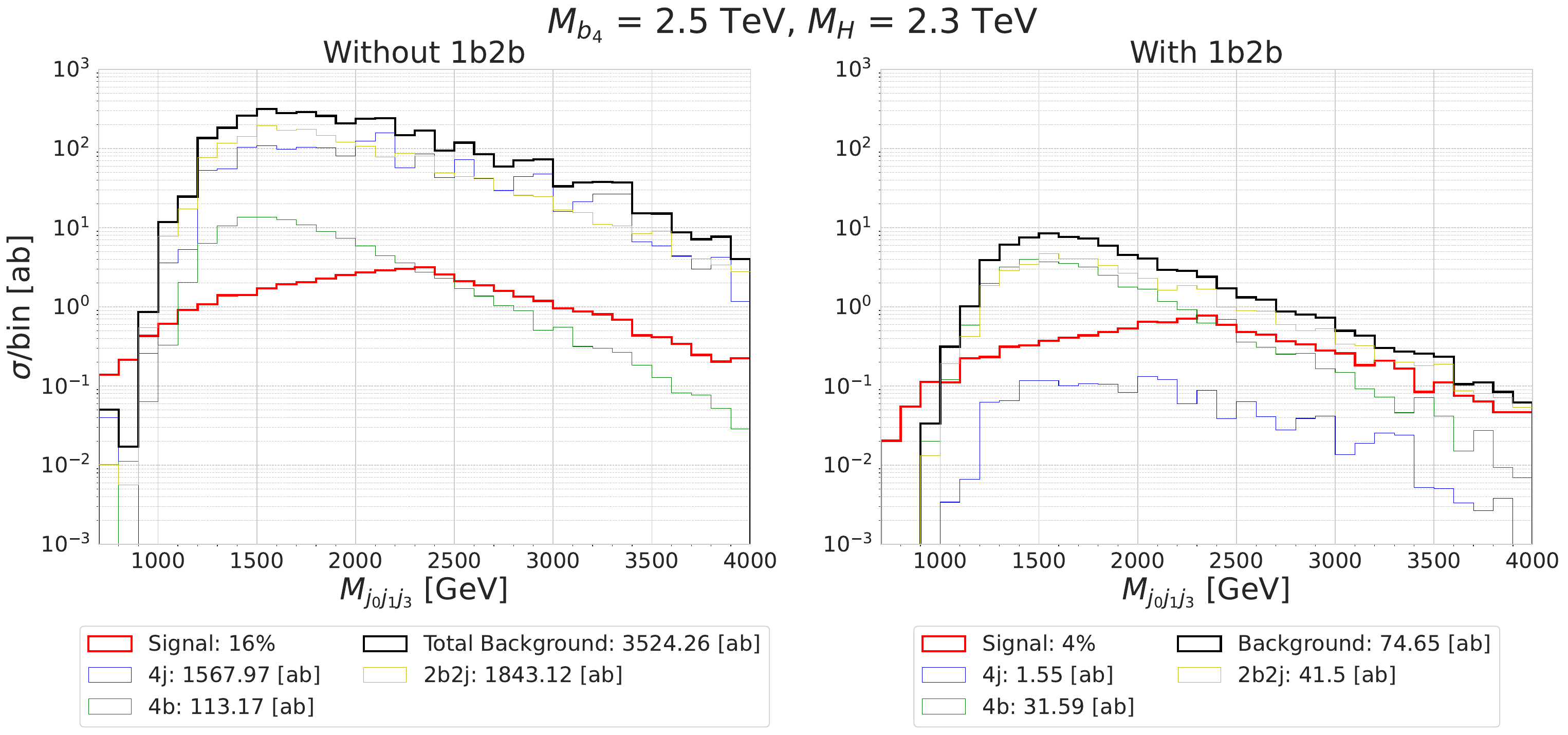}
\end{center}
\caption{$pp\to 6b$: di-jet and tri-jet invariant mass distributions for various $b_4$ and $H$ mass configurations after all cuts from table~\ref{tab:6bCuts} have been applied. Left panels show the results prior to the application of the 1b2b tagger.}
\label{fig:MJJJ2}
\end{figure}

\begin{figure}[h]
  \begin{center}
    \includegraphics[width=0.75 \linewidth]{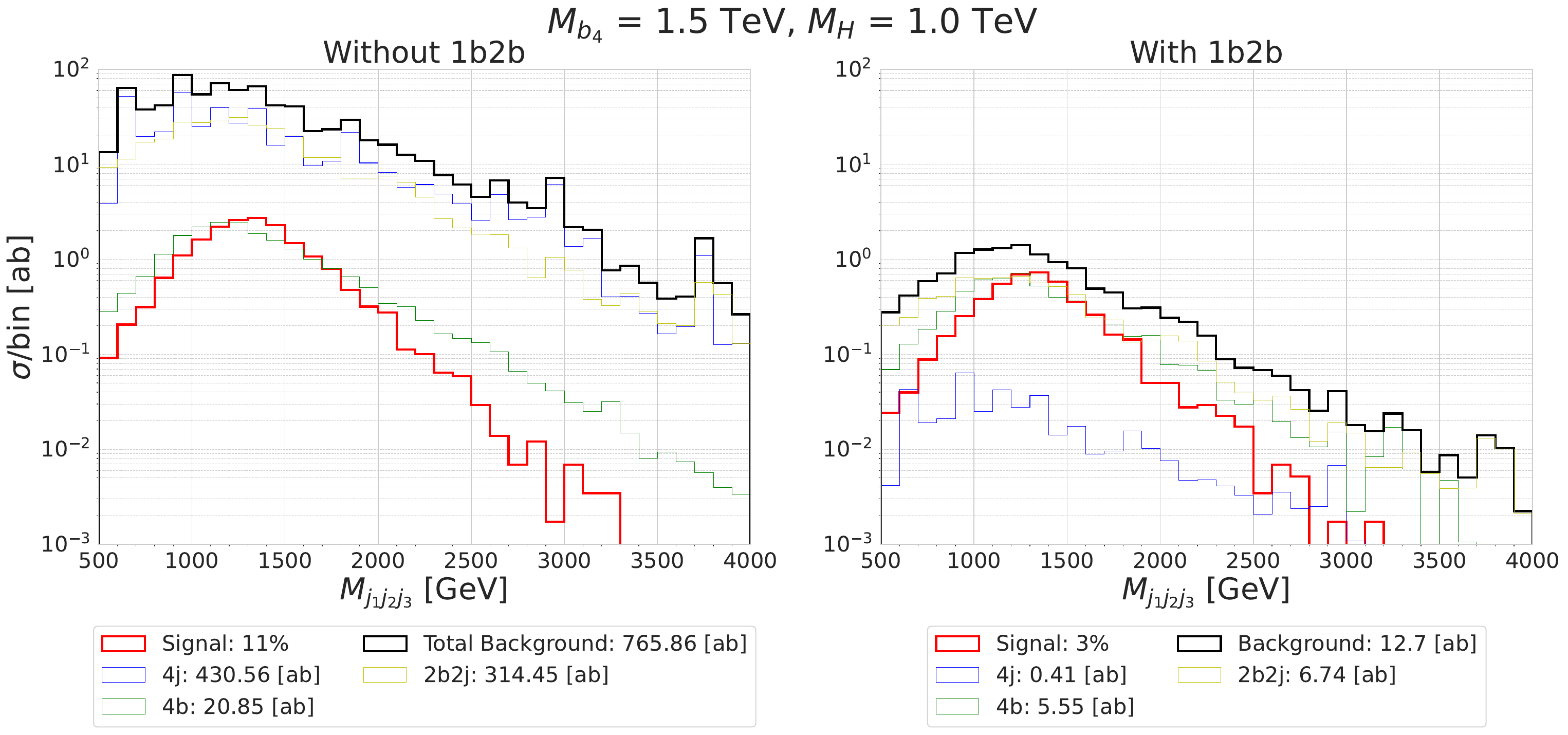}
    \includegraphics[width=0.75 \linewidth]{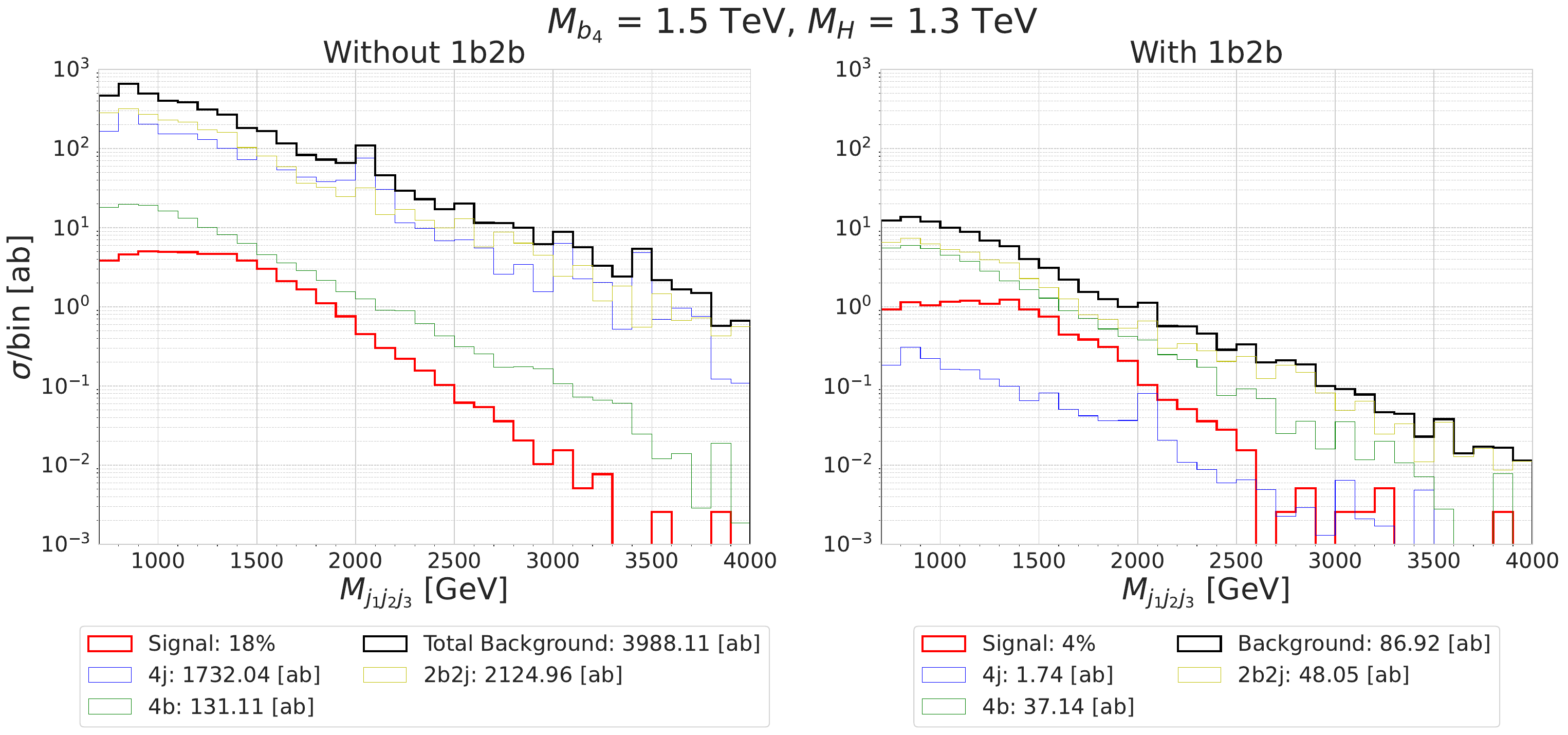}
    \includegraphics[width=0.75 \linewidth]{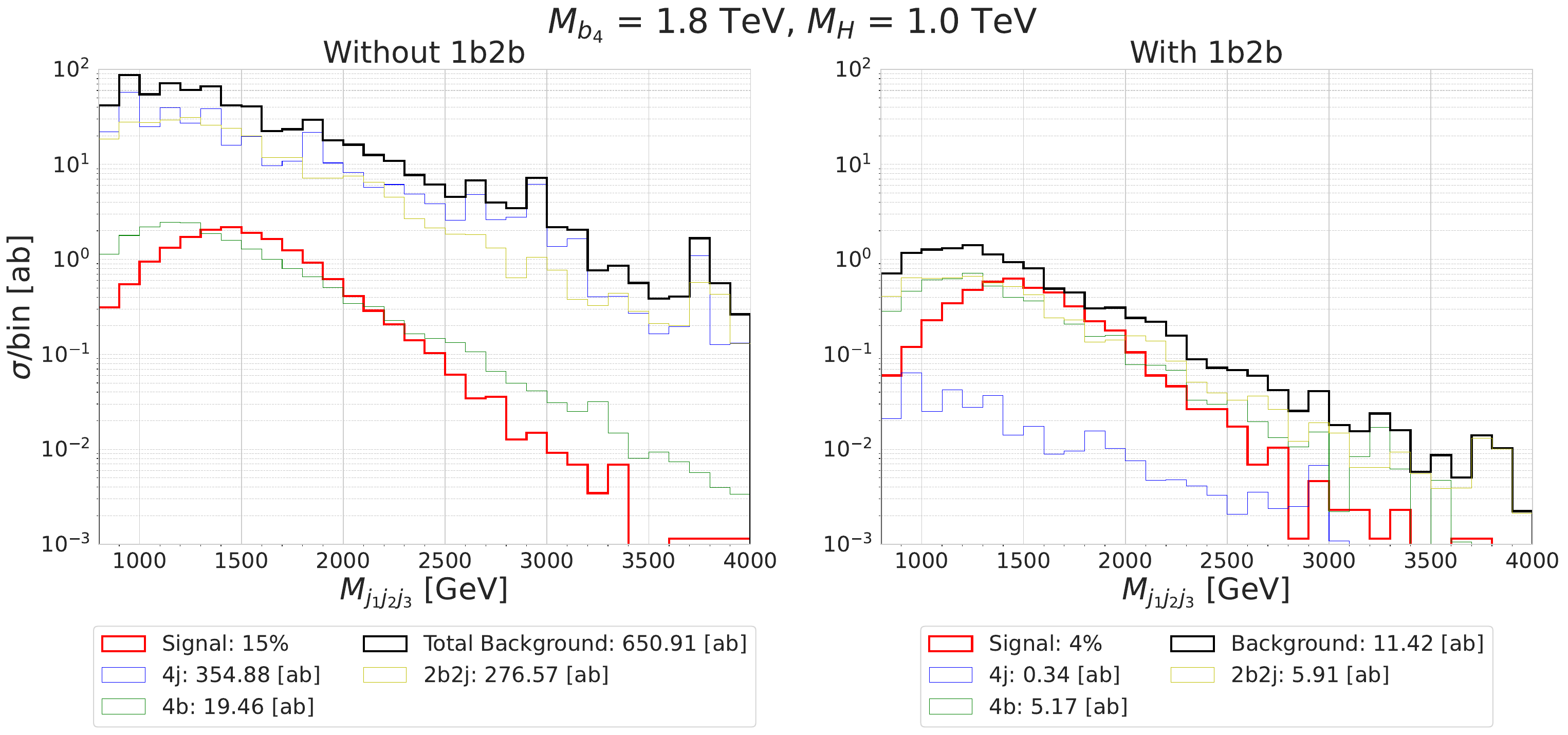}
    \includegraphics[width=0.75 \linewidth]{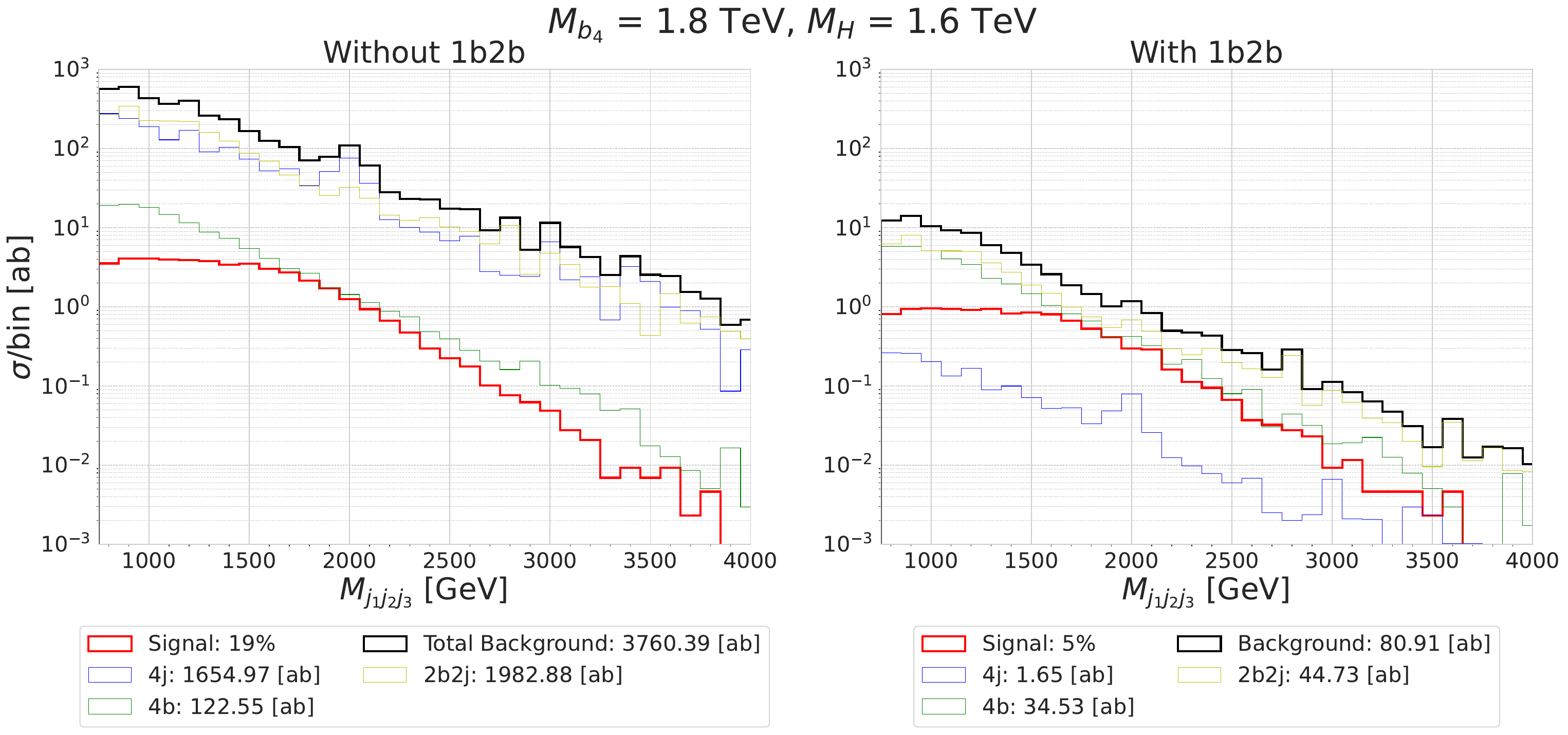}
  \end{center}
  \caption{$pp\to 6b$: tri-jet invariant mass distributions for various $b_4$ and $H$ mass configurations after all cuts from table~\ref{tab:6bCuts} have been applied. Left panels show the results prior to the application of the 1b2b tagger. }
  \label{fig:MJ123_1}
  \end{figure}
  
  \begin{figure}[h]
  \begin{center}
    \includegraphics[width=0.75 \linewidth]{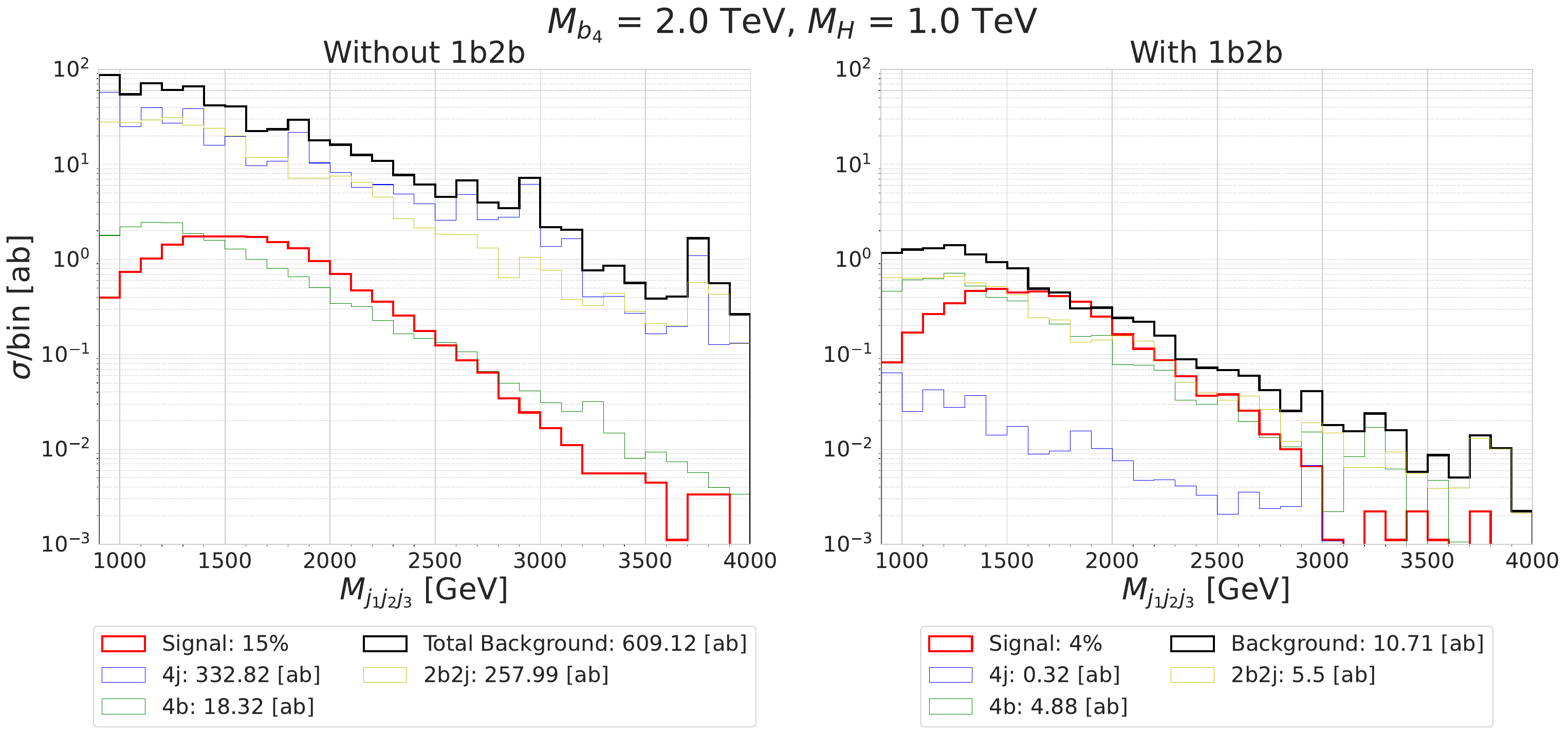}
    \includegraphics[width=0.75 \linewidth]{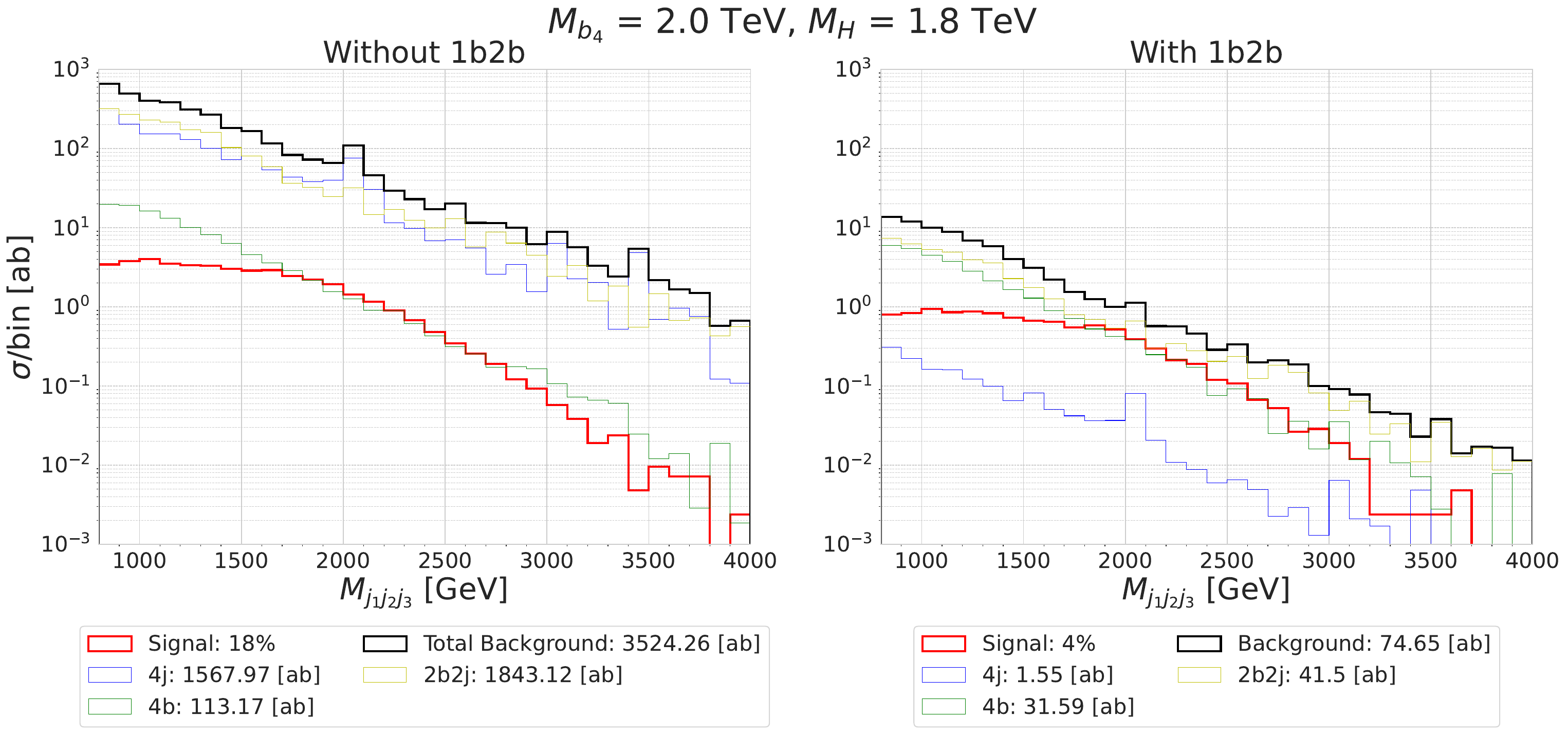}
    \includegraphics[width=0.75 \linewidth]{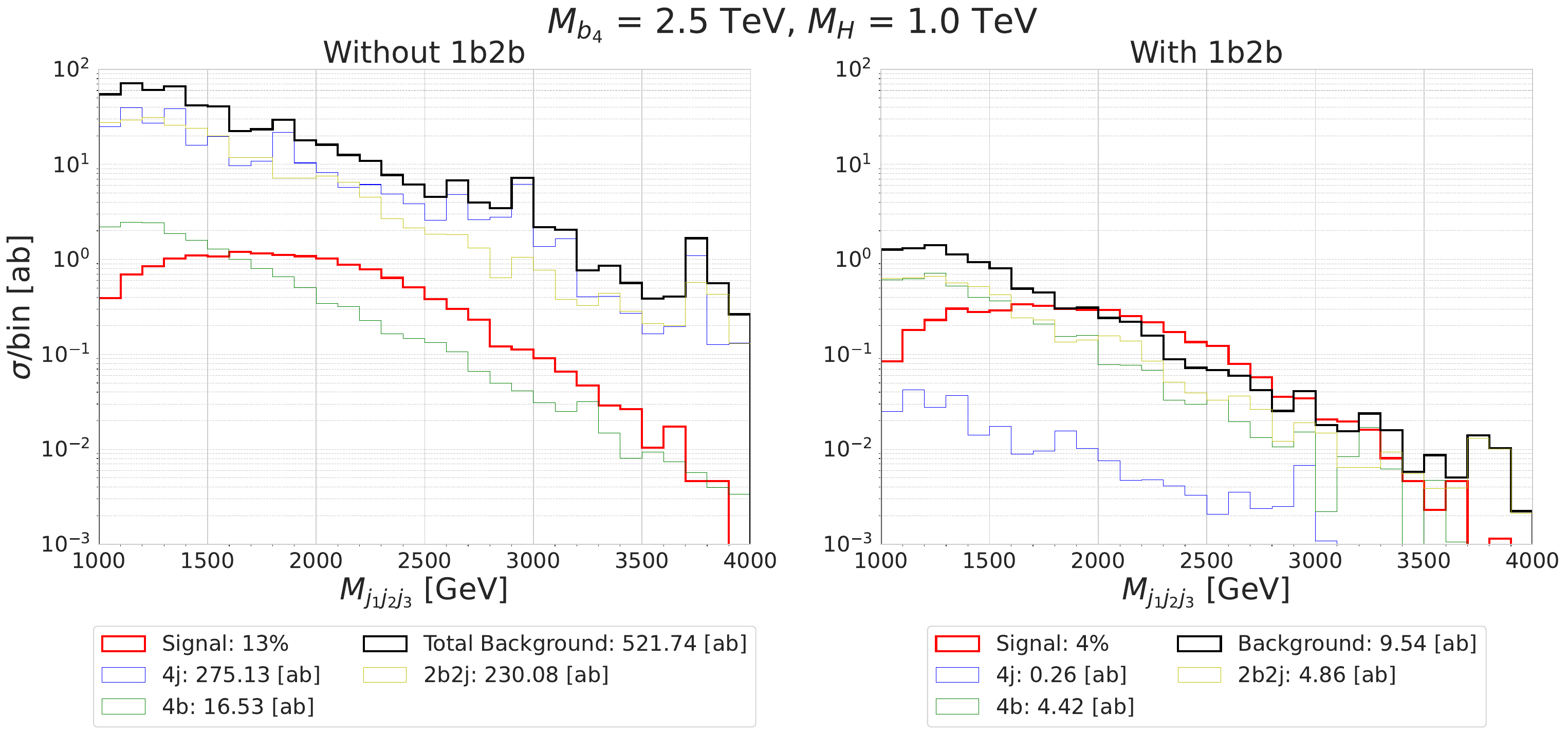}
    \includegraphics[width=0.75 \linewidth]{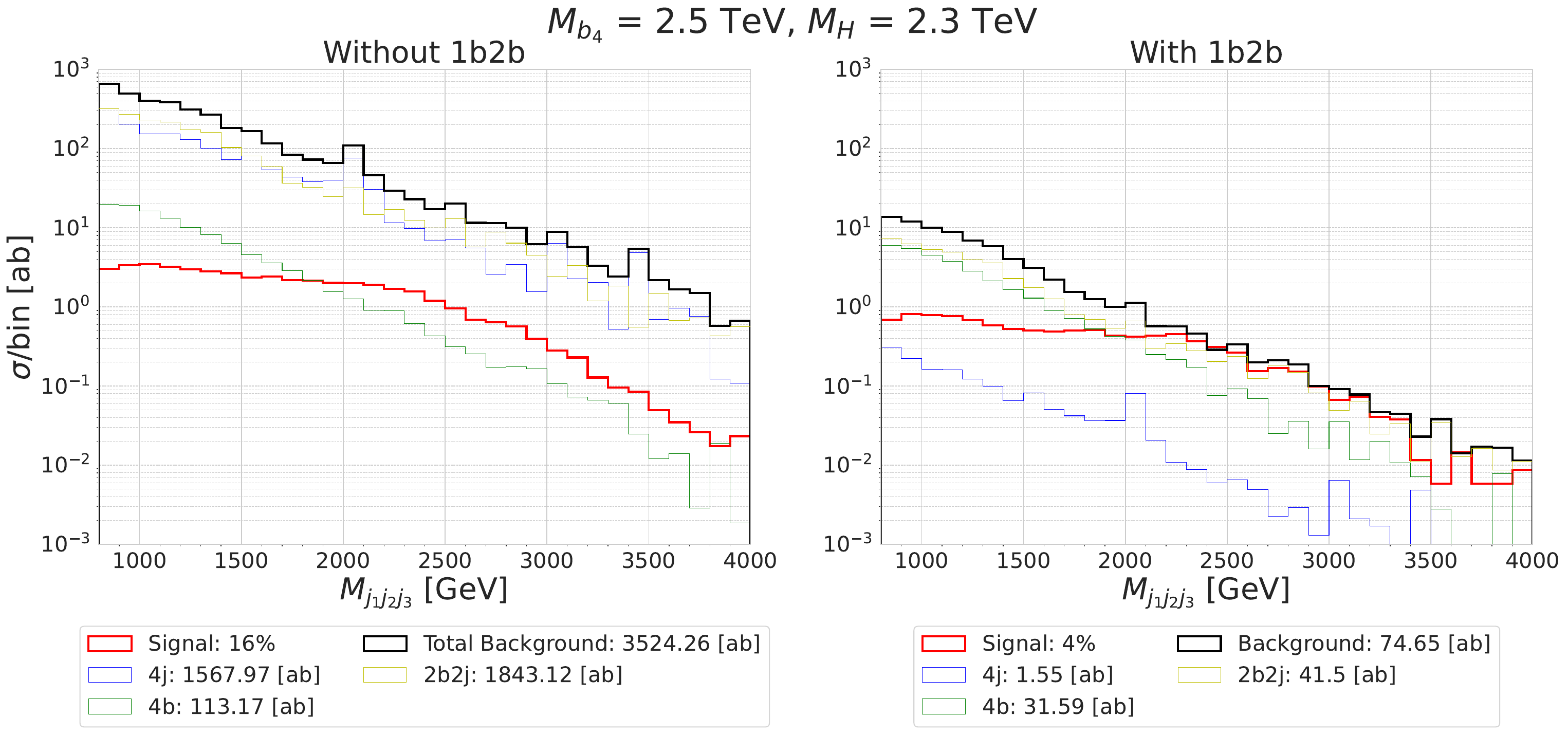}
  \end{center}
  \caption{$pp\to 6b$: tri-jet invariant mass distributions for various $b_4$ and $H$ mass configurations after all cuts from table~\ref{tab:6bCuts} have been applied. Left panels show the results prior to the application of the 1b2b tagger.}
  \label{fig:MJ123_2}
  \end{figure}


\begin{figure}[ht]
  \begin{center}
    \includegraphics[width=0.77 \linewidth]{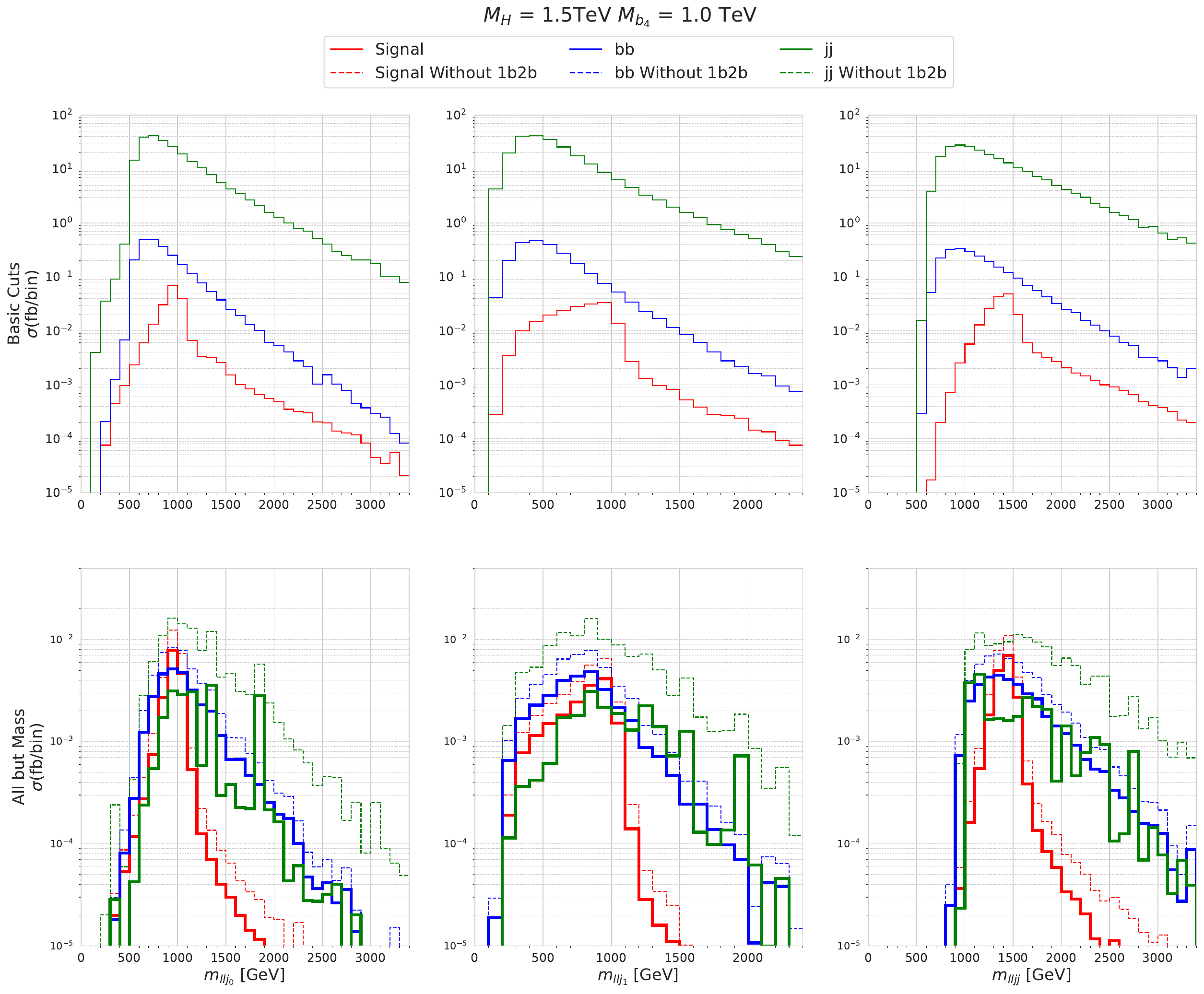}
  \end{center}
  \caption{$pp\to  2\ell 2b$:  Invariant mass of the leading jet combined with the two leptons.}
  \label{fig:MZbb_1510}
\end{figure}

\begin{figure}[ht]
  \begin{center}
    \includegraphics[width=0.77 \linewidth]{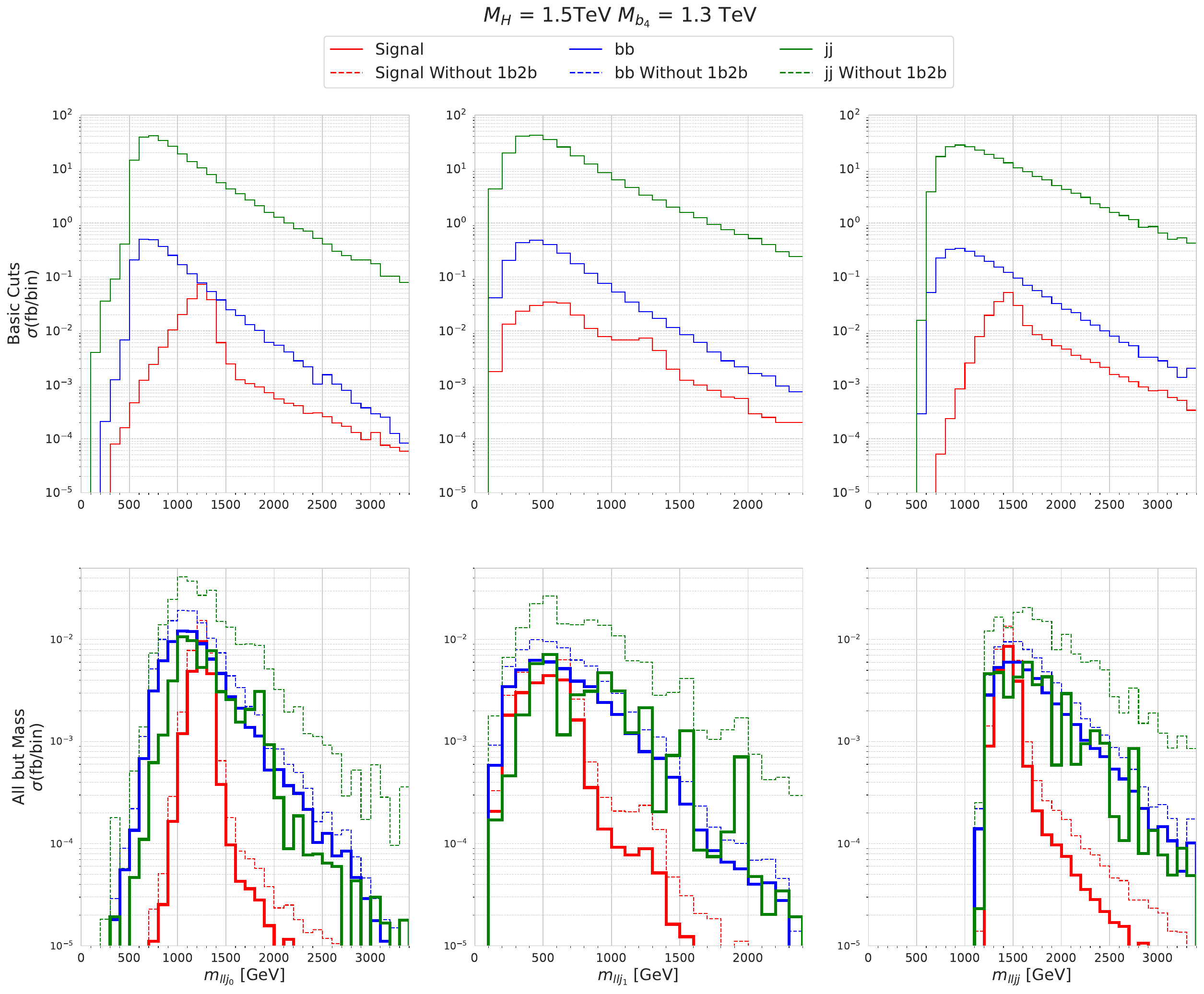}
  \end{center}
  \caption{  $pp\to  2\ell 2b$: invariant mass of the leading jet combined with the two leptons.}
  \label{fig:MZbb_1513}
\end{figure}

\begin{figure}[ht]: inv
  \begin{center}
    \includegraphics[width=0.77 \linewidth]{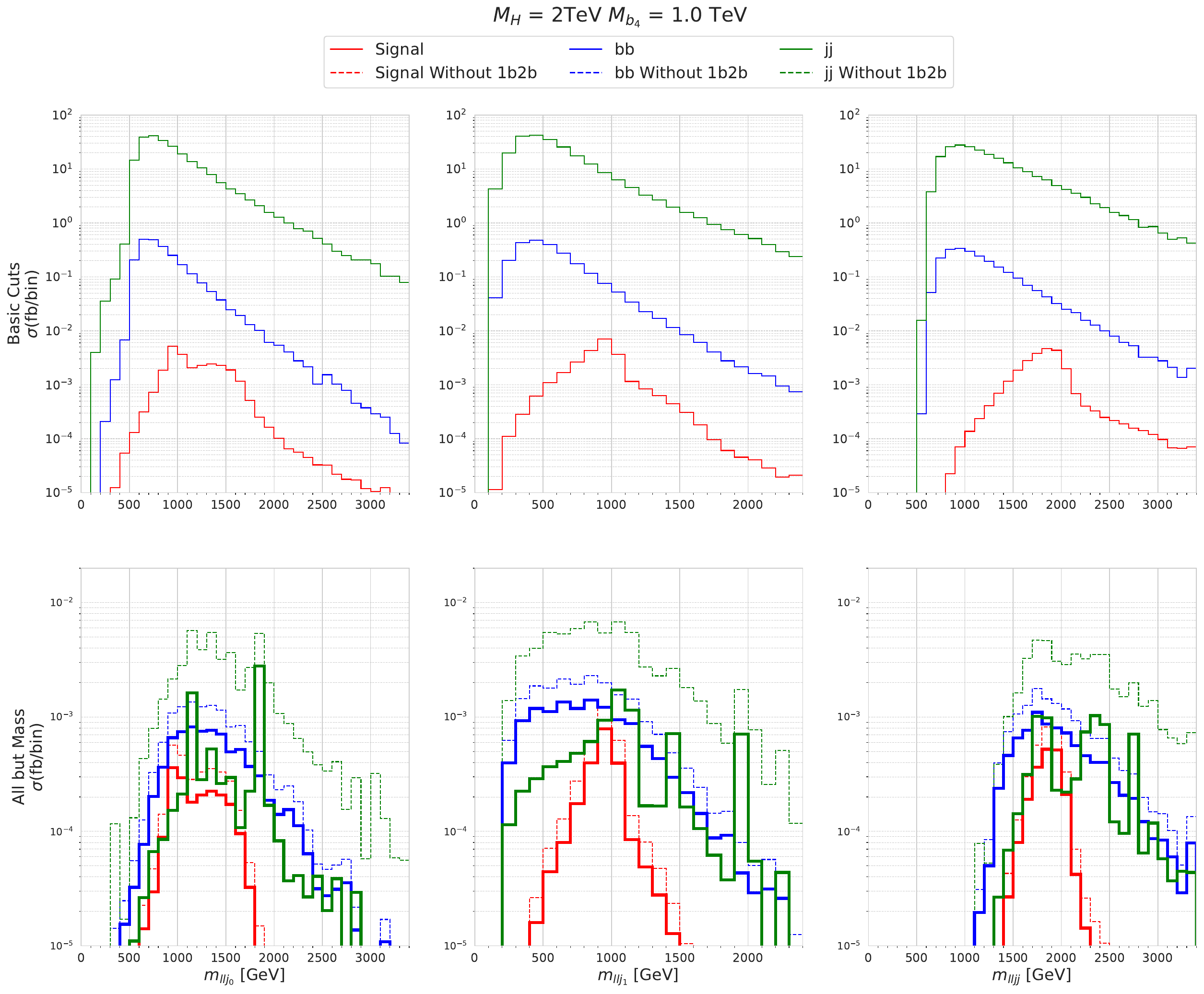}
  \end{center}
  \caption{$pp\to  2\ell 2b$: invariant mass of the leading jet combined with the two leptons.}
  \label{fig:MZbb_2010}
\end{figure}

\begin{figure}[ht]
  \begin{center}
    \includegraphics[width=0.77 \linewidth]{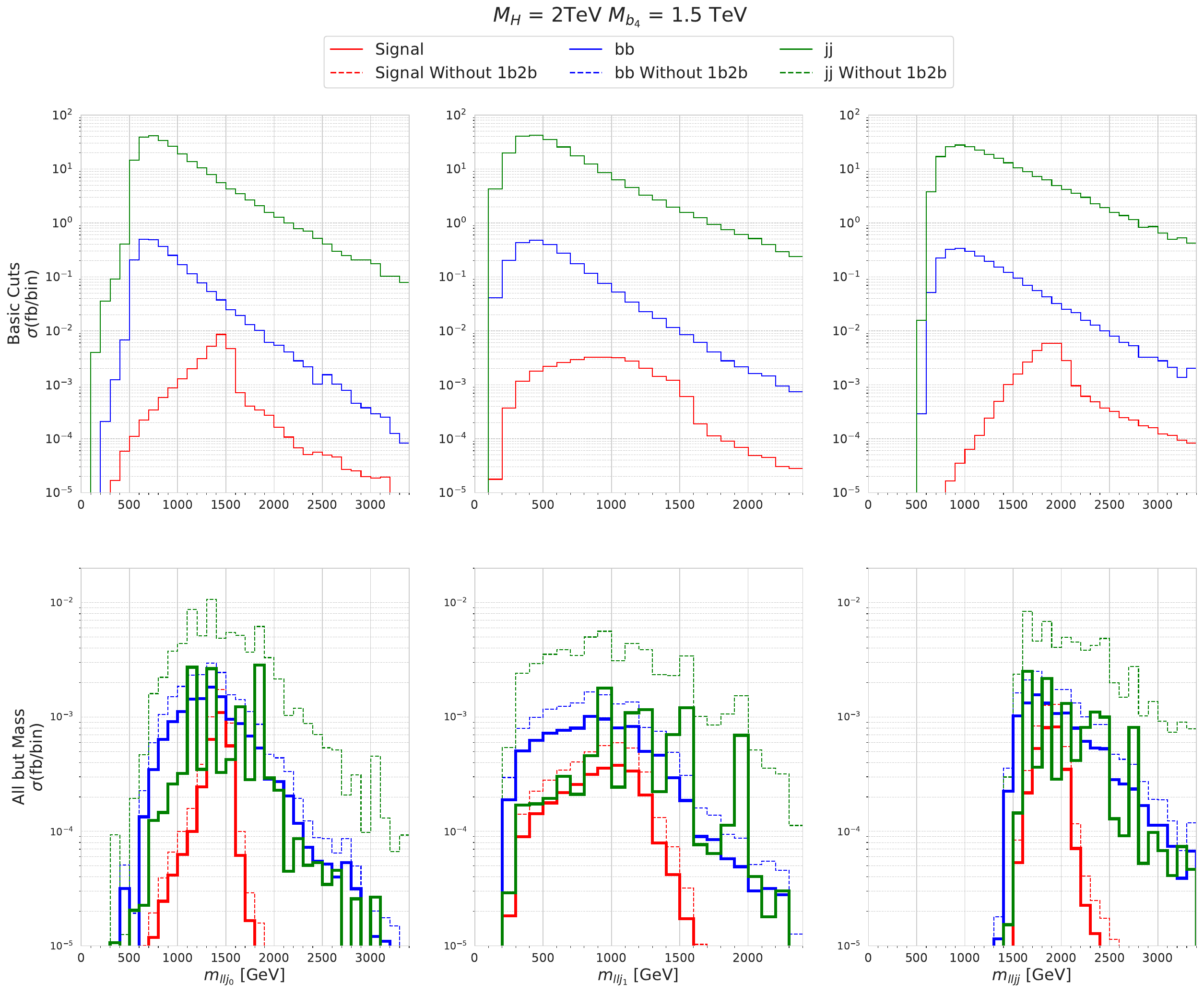}
  \end{center}
  \caption{$pp\to  2\ell 2b$: invariant mass of the leading jet combined with the two leptons.}
  \label{fig:MZbb_2015}
\end{figure}

\begin{figure}[ht]
  \begin{center}
    \includegraphics[width=0.77 \linewidth]{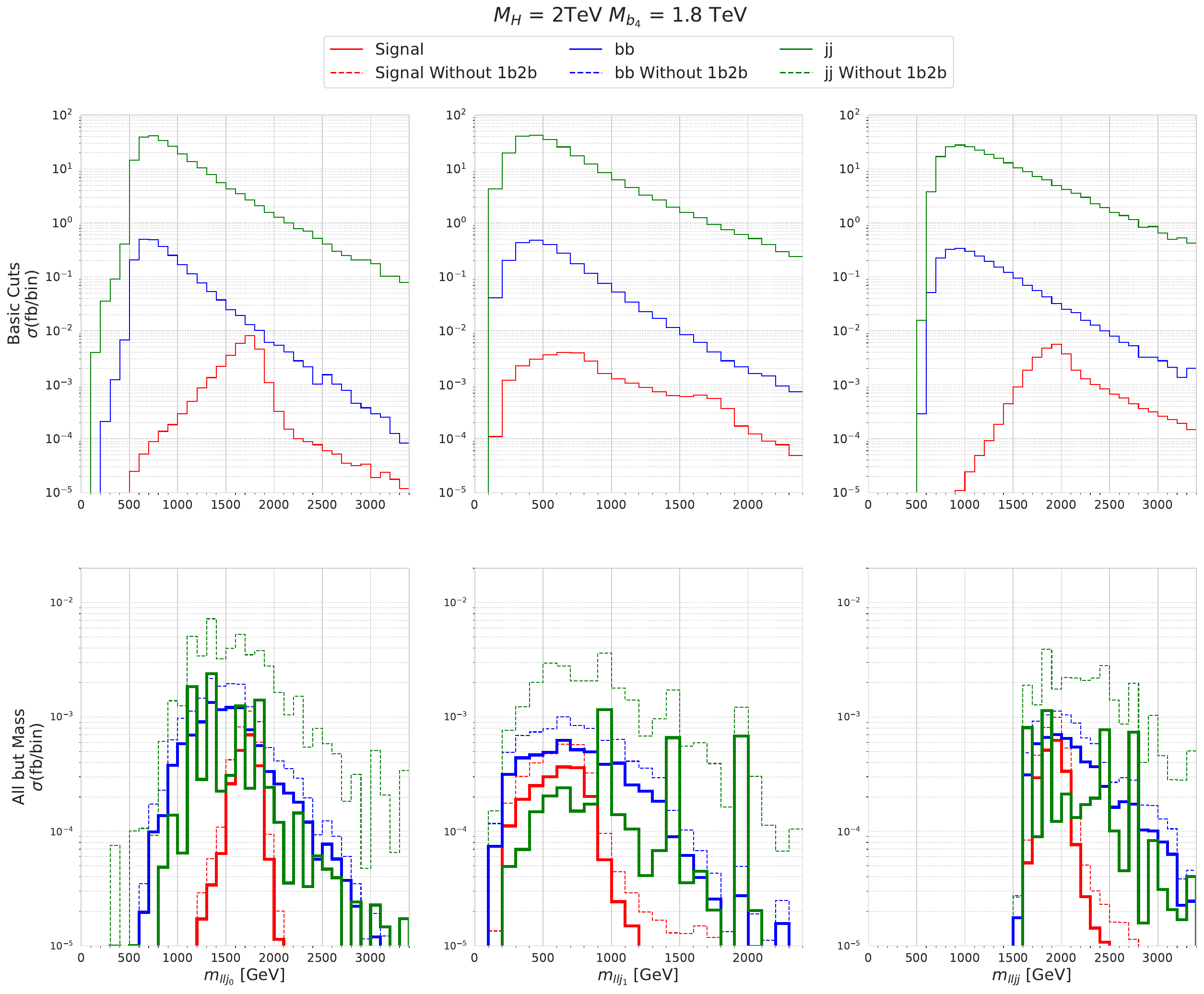}
  \end{center}
  \caption{$pp\to  2\ell 2b$: invariant mass of the leading jet combined with the two leptons.}
  \label{fig:MZbb_2018}
\end{figure}

\begin{figure}[ht]
  \begin{center}
    \includegraphics[width=0.77 \linewidth]{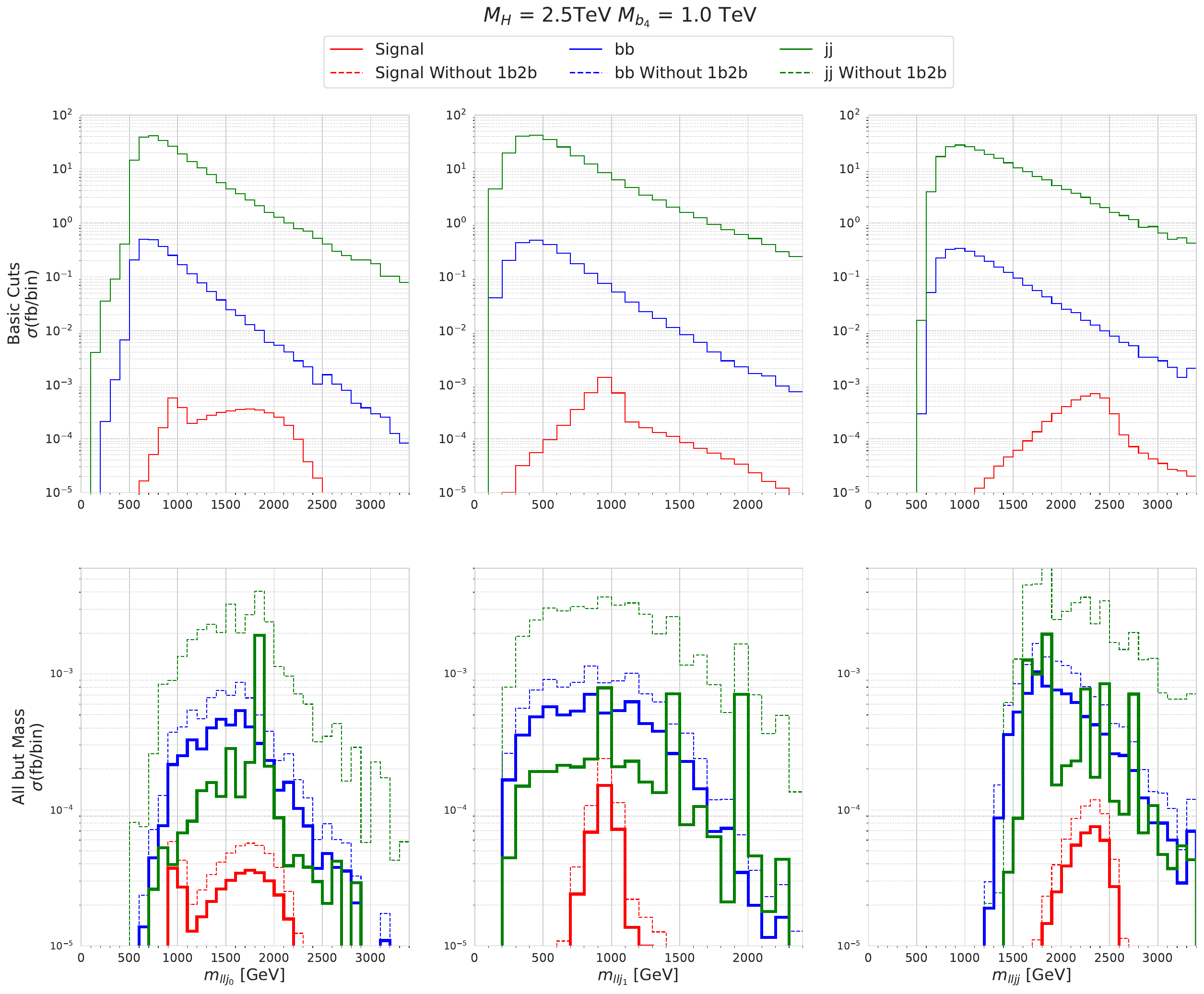}
  \end{center}
  \caption{$pp\to  2\ell 2b$: invariant mass of the leading jet combined with the two leptons.}
  \label{fig:MZbb_2510}
\end{figure}

\begin{figure}[ht]
  \begin{center}
    \includegraphics[width=0.77 \linewidth]{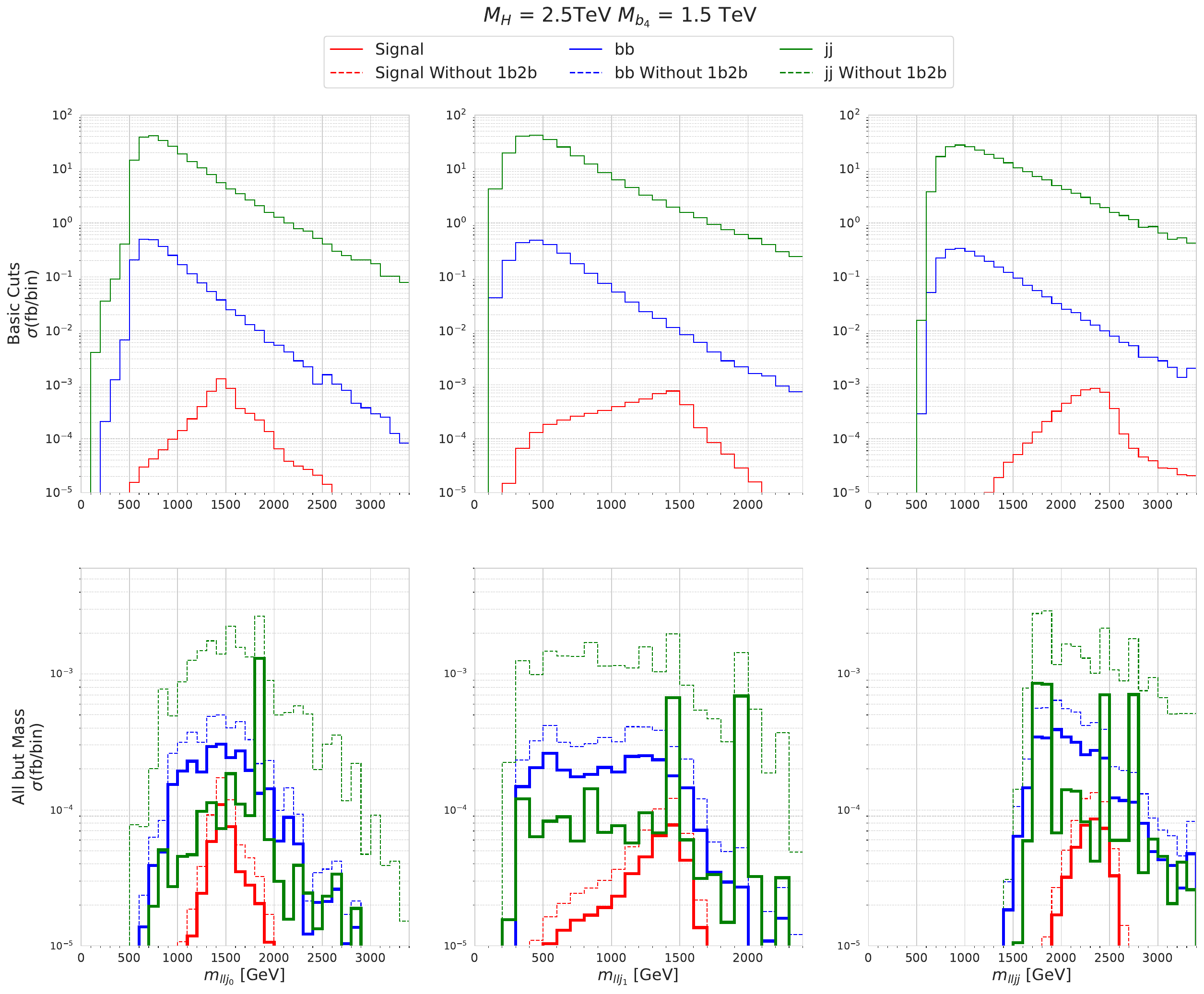}
  \end{center}
  \caption{$pp\to  2\ell 2b$: invariant mass of the leading jet combined with the two leptons.}
  \label{fig:MZbb_2515}
\end{figure}

\begin{figure}[ht]
  \begin{center}
    \includegraphics[width=0.77 \linewidth]{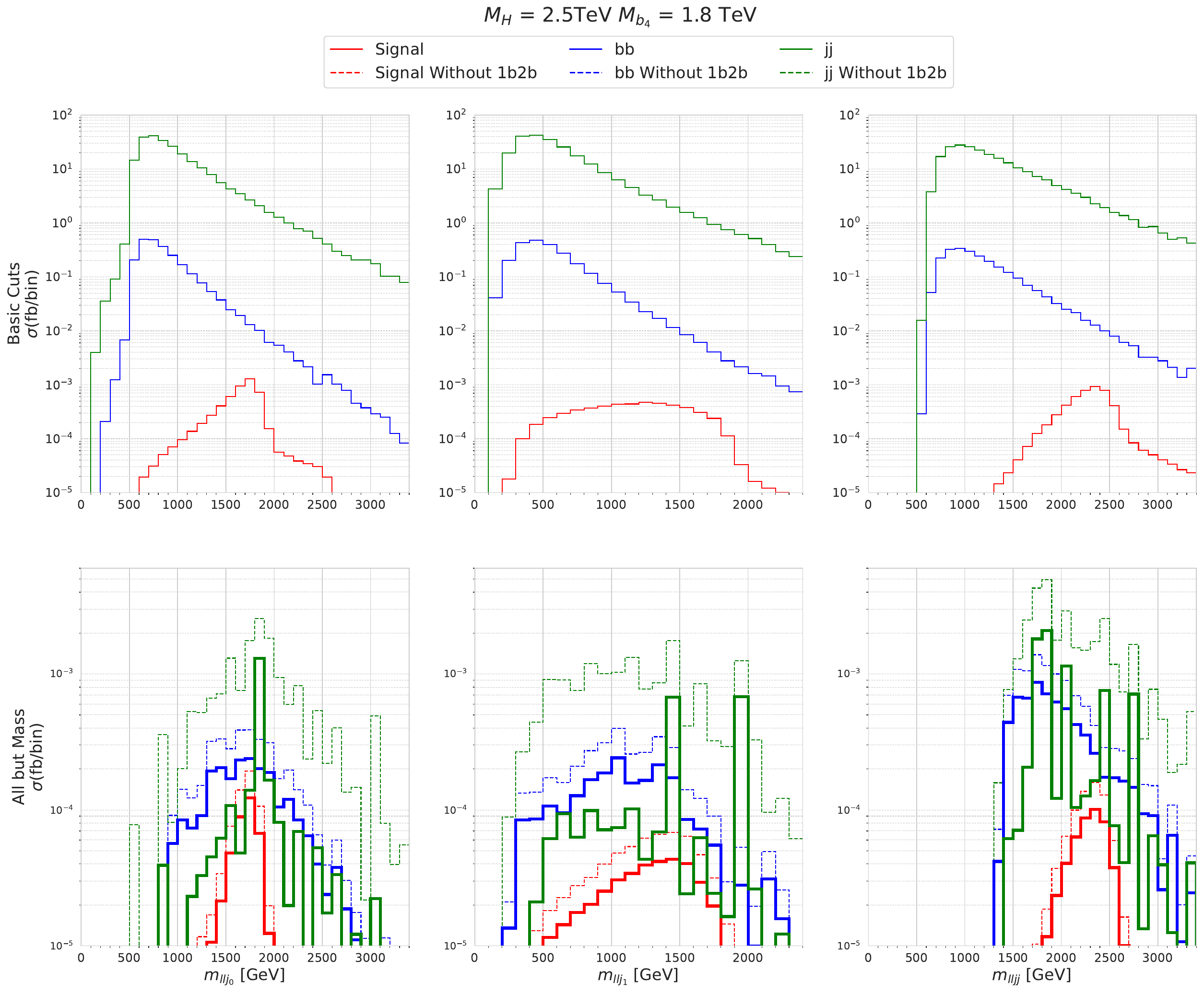}
  \end{center}
  \caption{$pp\to  2\ell 2b$: invariant mass of the leading jet combined with the two leptons.}
  \label{fig:MZbb_2518}
\end{figure}

\begin{figure}[ht]
  \begin{center}
    \includegraphics[width=0.77 \linewidth]{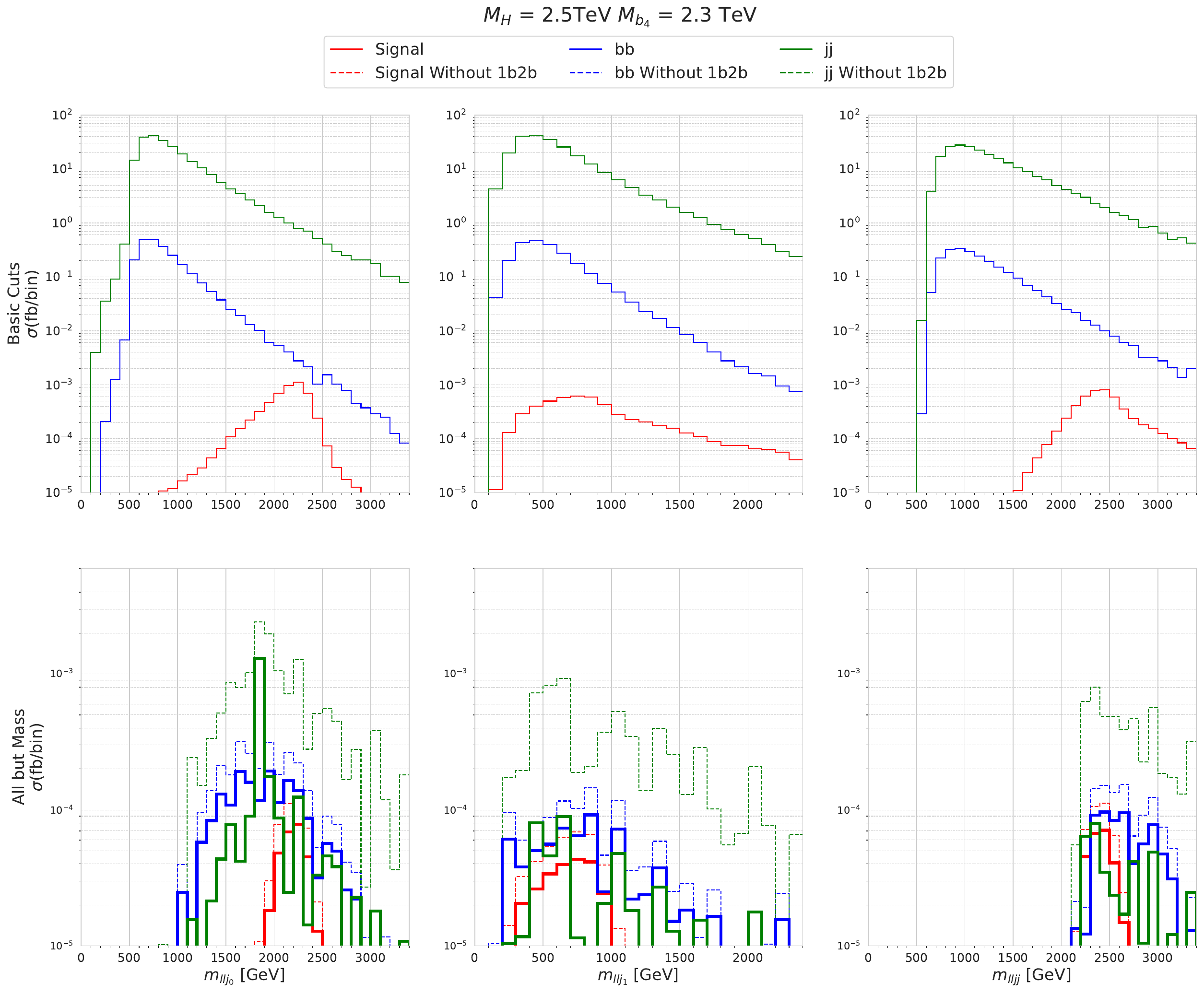}
      \end{center}
  \caption{$pp\to  2\ell 2b$: invariant mass of the leading jet combined with the two leptons.}
  \label{fig:MZbb_2523}
\end{figure}

\bibliography{references}{}
\bibliographystyle{JHEP} 

\end{document}